\documentclass{article}
\usepackage{lmodern}
\usepackage[utf8]{inputenc}
\usepackage[T1]{fontenc}
\usepackage[english]{babel}

\usepackage[a4paper, margin=1in]{geometry}

\usepackage{mleftright,xparse,xstring}
\usepackage[parfill]{parskip}
\usepackage{todonotes}
\usepackage{stmaryrd}
\usepackage{here}

\usepackage[autostyle]{csquotes}
\usepackage[]{hyperref}
\usepackage{xcolor}
\hypersetup{
    colorlinks,
    linkcolor={red!50!black},
    citecolor={blue!50!black},
    urlcolor={blue!80!black}
}

\usepackage{subcaption}
\usepackage{enumitem}

\usepackage{tikz}
\usetikzlibrary{arrows,positioning,intersections, decorations.pathreplacing}
\usetikzlibrary{backgrounds}
\usetikzlibrary{trees}

\tikzstyle{hvertex}=[thick,circle,inner sep=0.cm, minimum size=2.5mm, fill=white, draw=black]
\tikzstyle{hedge}=[ultra thick]
\tikzstyle{harc}=[ultra thick, ->]
\tikzstyle{point}=[draw,circle,inner sep=0.cm, minimum size=1mm, fill=black]
\tikzstyle{face}=[color=auchblau,fill=hellblau,thick]
\tikzstyle{nface}=[color=hellblau,fill=hellblau,thick] 
\usetikzlibrary{arrows.meta,patterns}
\usetikzlibrary{ipe}

\usepackage{graphicx}

\usepackage{tikz}
\usepackage{pgfplots}
\usetikzlibrary{arrows,positioning,decorations.pathreplacing}
\usetikzlibrary{calc,shapes,chains}
\usetikzlibrary{arrows.meta}
\tikzset{%
  every picture/.style={scale=1.0},%
  arc/.style={->, >=latex},%
  vertex/.style={draw=black,circle,fill=black,text width=3pt,inner sep=0mm,outer sep=1pt},%
  every label/.append style={rectangle}%
}

\def\tikzAngleOfLine(#1)(#2)#3{%
    \pgfmathanglebetweenpoints{\pgfpointanchor{#1}{center}}{\pgfpointanchor{#2}{center}}%
    \pgfmathsetmacro{#3}{\pgfmathresult}%
}

\tikzstyle{ipe stylesheet} = [
  ipe import,
  even odd rule,
  line join=round,
  line cap=butt,
  ipe pen normal/.style={line width=0.4},
  ipe pen heavier/.style={line width=0.8},
  ipe pen fat/.style={line width=1.2},
  ipe pen ultrafat/.style={line width=2},
  ipe pen normal,
  ipe mark normal/.style={ipe mark scale=3},
  ipe mark large/.style={ipe mark scale=5},
  ipe mark small/.style={ipe mark scale=2},
  ipe mark tiny/.style={ipe mark scale=1.1},
  ipe mark normal,
  /pgf/arrow keys/.cd,
  ipe arrow normal/.style={scale=7},
  ipe arrow large/.style={scale=10},
  ipe arrow small/.style={scale=5},
  ipe arrow tiny/.style={scale=3},
  ipe arrow normal,
  /tikz/.cd,
  ipe arrows, 
  <->/.tip = latex',
  ipe dash normal/.style={dash pattern=},
  ipe dash dashed/.style={dash pattern=on 4bp off 4bp},
  ipe dash dotted/.style={dash pattern=on 1bp off 3bp},
  ipe dash dash dotted/.style={dash pattern=on 4bp off 2bp on 1bp off 2bp},
  ipe dash dash dot dotted/.style={dash pattern=on 4bp off 2bp on 1bp off 2bp on 1bp off 2bp},
  ipe dash normal,
  ipe node/.append style={font=\normalsize},
  ipe stretch normal/.style={ipe node stretch=1},
  ipe stretch normal,
  ipe opacity 10/.style={opacity=0.1},
  ipe opacity 30/.style={opacity=0.3},
  ipe opacity 50/.style={opacity=0.5},
  ipe opacity 75/.style={opacity=0.75},
  ipe opacity opaque/.style={opacity=1},
  ipe opacity opaque,
]

\definecolor{red}{rgb}{1,0,0}
\definecolor{green}{rgb}{0,1,0}
\definecolor{blue}{rgb}{0,0,1}
\definecolor{yellow}{rgb}{1,1,0}
\definecolor{orange}{rgb}{1,0.647,0}
\definecolor{gold}{rgb}{1,0.843,0}
\definecolor{purple}{rgb}{0.627,0.125,0.941}
\definecolor{gray}{rgb}{0.745,0.745,0.745}
\definecolor{brown}{rgb}{0.647,0.165,0.165}
\definecolor{navy}{rgb}{0,0,0.502}
\definecolor{pink}{rgb}{1,0.753,0.796}
\definecolor{seagreen}{rgb}{0.18,0.545,0.341}
\definecolor{turquoise}{rgb}{0.251,0.878,0.816}
\definecolor{violet}{rgb}{0.933,0.51,0.933}
\definecolor{darkblue}{rgb}{0,0,0.545}
\definecolor{darkcyan}{rgb}{0,0.545,0.545}
\definecolor{darkgray}{rgb}{0.663,0.663,0.663}
\definecolor{darkgreen}{rgb}{0,0.392,0}
\definecolor{darkmagenta}{rgb}{0.545,0,0.545}
\definecolor{darkorange}{rgb}{1,0.549,0}
\definecolor{darkred}{rgb}{0.545,0,0}
\definecolor{lightblue}{rgb}{0.678,0.847,0.902}
\definecolor{lightcyan}{rgb}{0.878,1,1}
\definecolor{lightgray}{rgb}{0.827,0.827,0.827}
\definecolor{lightgreen}{rgb}{0.565,0.933,0.565}
\definecolor{lightyellow}{rgb}{1,1,0.878}
\definecolor{black}{rgb}{0,0,0}
\definecolor{white}{rgb}{1,1,1}

\usepackage[linesnumbered,ruled]{algorithm2e}
\usepackage{setspace}

\SetKwInput{KwParameter}{Parameter}
\SetKwInput{KwInput}{Input}
\SetKwInput{KwOutput}{Output}
\SetKwInOut{Task}{Task}
\SetKwRepeat{Do}{do}{while}%

\usepackage{amsmath}
\usepackage{amssymb}
\usepackage{dsfont}
\usepackage{mathtools}
\usepackage{accents}

\newtoggle{clara}

\togglefalse{clara}

\iftoggle{clara}{
	\usepackage[nameinlink]{cleveref}
	\usepackage{amsthm}
}{
	\usepackage{amsthm}
	\usepackage[nameinlink]{cleveref}
}

\newtheoremstyle{break}
  {\parskip}{\parskip}%
  {\itshape}{}%
  {\bfseries}{}%
  {\newline}{}%

\newtheoremstyle{named}
  {\parskip}{\parskip}%
  {\itshape}{}%
  {\bfseries}{}%
  {\newline}{\thmnote{#3 }#1}%

\theoremstyle{break}

\newtheorem{definition}{Definition}
\newtheorem{lemma}[definition]{Lemma}
\newtheorem{theorem}[definition]{Theorem}

\newtheorem{corollary}[definition]{Corollary}
\newtheorem{observation}[definition]{Observation}

\theoremstyle{named}

\Crefname{definition}{Definition}{Definitions}
\crefname{definition}{definition}{definitions}
\Crefname{lemma}{Lemma}{Lemmas}
\crefname{lemma}{Lemma}{Lemmas}
\Crefname{theorem}{Theorem}{Theorems}
\crefname{theorem}{Theorem}{Theorems}
\Crefname{proposition}{Proposition}{Propositions}
\crefname{proposition}{Proposition}{Propositions}
\Crefname{corollary}{Corollary}{Corollaries}
\crefname{corollary}{Corollary}{Corollaries}
\Crefname{observation}{Observation}{Observations}
\crefname{observation}{Observation}{Observations}
\Crefname{remark}{Remark}{Remarks}
\crefname{remark}{Remark}{Remarks}
\Crefname{algocf}{Algorithm}{Algorithms}
\crefname{algocf}{Algorithm}{Algorithms}
\Crefname{algorithm}{Algorithm}{Algorithms}
\crefname{algorithm}{Algorithm}{Algorithms}
\Crefname{problem}{Problem}{Problems}
\crefname{problem}{Problem}{Problems}
\Crefname{figure}{Figure}{Figures}
\crefname{figure}{Figure}{Figures}

\newlist{definitionenum}{enumerate}{1}
\setlist[definitionenum]{label=(\roman*), ref={\thedefinition~(\roman*)}}
\crefalias{definitionenumi}{definition}

\newlist{lemmaenum}{enumerate}{1}
\setlist[lemmaenum]{label=(\roman*), ref={\thelemma~(\roman*)}}
\crefalias{lemmaenumi}{lemma}

\newlist{theoremenum}{enumerate}{1}
\setlist[theoremenum]{label=(\roman*), ref={\thetheorem~(\roman*)}}
\crefalias{theoremenumi}{theorem}

\newlist{propositionenum}{enumerate}{1}
\setlist[propositionenum]{label=(\roman*), ref={\theproposition~(\roman*)}}
\crefalias{propositionenumi}{proposition}

\newlist{corollaryenum}{enumerate}{1}
\setlist[corollaryenum]{label=(\roman*), ref={\thecorollary~(\roman*)}}
\crefalias{corollaryenumi}{corollary}

\newlist{observationenum}{enumerate}{1}
\setlist[observationenum]{label=(\roman*), ref={\theobservation~(\roman*)}}
\crefalias{observationenumi}{observation}

\newlist{remarkenum}{enumerate}{1}
\setlist[remarkenum]{label=(\roman*), ref={\theremark~(\roman*)}}
\crefalias{remarkenumi}{remark}

\makeatletter

\providecommand*{\cupdot}{%
  \mathbin{%
    \mathpalette\@cupdot{}%
  }%
}
\newcommand*{\@cupdot}[2]{%
  \ooalign{%
    $\m@th#1\cup$\cr
    \sbox0{$#1\cup$}%
    \dimen@=\ht0 %
    \sbox0{$\m@th#1\cdot$}%
    \advance\dimen@ by -\ht0 %
    \dimen@=.5\dimen@
    \hidewidth\raise\dimen@\box0\hidewidth
  }%
}

\providecommand*{\bigcupdot}{%
  \mathop{%
    \vphantom{\bigcup}%
    \mathpalette\@bigcupdot{}%
  }%
}
\newcommand*{\@bigcupdot}[2]{%
  \ooalign{%
    $\m@th#1\bigcup$\cr
    \sbox0{$#1\bigcup$}%
    \dimen@=\ht0 %
    \advance\dimen@ by -\dp0 %
    \sbox0{\scalebox{1.5}{$\m@th#1\cdot$}}%
    \advance\dimen@ by -\ht0 %
    \dimen@=.5\dimen@
    \hidewidth\raise\dimen@\box0\hidewidth
  }%
}

\def\mathcalfactory#1{%
	\expandafter\newcommand\csname #1cal\endcsname{\mathcal{#1}}%
}

\def\mathbbfactory#1{%
	\expandafter\newcommand\csname #1bb\endcsname{\mathbb{#1}}%
}

\newcounter{ctr}
\loop
	\stepcounter{ctr}
	\edef\X{\@Alph\c@ctr}
	\expandafter\mathbbfactory\X
	\expandafter\mathcalfactory\X
\ifnum\thectr<26
\repeat

\newcommand\Rbbp{\Rbb_{\ge 0}}

\newcommand\NPhard{NP-hard}

\DeclareFontFamily{U}{mathb}{\hyphenchar\font45}
\DeclareFontShape{U}{mathb}{m}{n}{
<-6> mathb5 <6-7> mathb6 <7-8> mathb7
<8-9> mathb8 <9-10> mathb9
<10-12> mathb10 <12-> mathb12
}{}
\DeclareSymbolFont{mathb}{U}{mathb}{m}{n}
\DeclareMathSymbol{\llcurly}{\mathrel}{mathb}{"CE}
\DeclareMathSymbol{\ggcurly}{\mathrel}{mathb}{"CF}

\NewDocumentCommand\xDeclarePairedDelimiter{mmm}
 {%
  \NewDocumentCommand#1{som}{%
   \IfNoValueTF{##2}
    {\IfBooleanTF{##1}{#2##3#3}{\mleft#2##3\mright#3}}
    {\mathopen{##2#2}##3\mathclose{##2#3}}%
  }%
 }

\xDeclarePairedDelimiter\floor{\lfloor}{\rfloor}
\xDeclarePairedDelimiter\ceil{\lceil}{\rceil}
\xDeclarePairedDelimiter\angles{\langle}{\rangle}
\xDeclarePairedDelimiter\braces{(}{)}
\xDeclarePairedDelimiter\set{\lbrace}{\rbrace}
\xDeclarePairedDelimiter\abs{|}{|}
\xDeclarePairedDelimiter\card{|}{|}
\xDeclarePairedDelimiter\norm{\|}{\|}

\NewDocumentCommand\seq{somm}{%
 \IfNoValueTF{#2}
  {\IfBooleanTF{#1}{(#3)_{#4}}{\mleft(#3\mright)_{#4}}}
  {\mathopen{#2(}#3\mathclose{#2)_{#4}}}%
}

\NewDocumentCommand\inv{som}{%
 \IfNoValueTF{#2}
  {\IfBooleanTF{#1}{(#3)^{-1}}{\mleft(#3\mright)^{-1}}}
  {\mathopen{#2(}#3\mathclose{#2)^{-1}}}%
}

\NewDocumentEnvironment{program}{oo}{
    \NewDocumentCommand\objective{mm}{##1 && ##2 \IfNoValueTF{#1}{\nonumber}{\tag{#1}} \IfNoValueF{#2}{\label{#2}} \span\\}
    
    \NewDocumentCommand\constraint{oom}{&& ##3 &&\IfNoValueF{##1}{\text{for all } \IfNoValueTF{##2}{##1}{##2}} \IfNoValueTF{##2}{\nonumber}{\label{##1}}}
    \align
}{
    \nonumber\endalign
}

\providecommand*{\sqsubseteqplus}{\mathbin{\mathpalette\@sqsubseteqplus{}}}

\newcommand*{\@sqsubseteqplus}[2]{%
  \ooalign{%
    $\m@th#1\sqsubseteq$\cr
    \sbox0{$#1\sqsubseteq$}%
    \dimen@=\ht0 %
    \sbox0{$\m@th#1+$}%
    \advance\dimen@ by -\ht0 %
    \dimen@=.5\dimen@
    \hidewidth\raise\dimen@\box0\hidewidth
  }%
}

\DeclareFontFamily{U}{matha}{\hyphenchar\font45}
\DeclareFontShape{U}{matha}{m}{n}{
      <5> <6> <7> <8> <9> <10> gen * matha
      <10.95> matha10 <12> <14.4> <17.28> <20.74> <24.88> matha12
      }{}
\DeclareSymbolFont{matha}{U}{matha}{m}{n}
\undef{\sqsubset}
\DeclareMathSymbol{\sqsubset}       {3}{mathb}{"80}
\undef{\sqsupset}
\DeclareMathSymbol{\sqsupset}       {3}{mathb}{"81}
\DeclareMathSymbol{\nsqsubset}      {3}{mathb}{"82}
\DeclareMathSymbol{\nsqsupset}      {3}{mathb}{"83}
\DeclareMathSymbol{\sqsubseteq}     {3}{mathb}{"84}
\DeclareMathSymbol{\sqsupseteq}     {3}{mathb}{"85}
\DeclareMathSymbol{\nsqsubseteq}    {3}{mathb}{"86}
\DeclareMathSymbol{\nsqsupseteq}    {3}{mathb}{"87}
\DeclareMathSymbol{\sqsubsetneq}    {3}{mathb}{"88}
\DeclareMathSymbol{\sqsupsetneq}    {3}{mathb}{"89}
\DeclareMathSymbol{\varsqsubsetneq} {3}{mathb}{"8A}
\DeclareMathSymbol{\varsqsupsetneq} {3}{mathb}{"8B}
\DeclareMathSymbol{\sqsubseteqq}    {3}{mathb}{"8C}
\DeclareMathSymbol{\sqsupseteqq}    {3}{mathb}{"8D}
\DeclareMathSymbol{\nsqsubseteqq}   {3}{mathb}{"8E}
\DeclareMathSymbol{\nsqsupseteqq}   {3}{mathb}{"8F}
\DeclareMathSymbol{\sqsubsetneqq}   {3}{mathb}{"90}
\DeclareMathSymbol{\sqsupsetneqq}   {3}{mathb}{"91}
\DeclareMathSymbol{\varsqsubsetneqq}{3}{mathb}{"92}
\DeclareMathSymbol{\varsqsupsetneqq}{3}{mathb}{"93}
\DeclareMathSymbol{\sqSubset}       {3}{mathb}{"94}
\DeclareMathSymbol{\sqSupset}       {3}{mathb}{"95}
\DeclareMathSymbol{\nsqSubset}      {3}{mathb}{"96}
\DeclareMathSymbol{\nsqSupset}      {3}{mathb}{"97}

\DeclareDocumentCommand\minbudget{}{\textsc{Budget Minimization
}}

\makeatother

\title{Budget Minimization with Precedence Constraints}
\author{Marinus Gottschau\thanks{
     Department of Mathematics and School of Management,
    Technische Universit\"{a}t M\"{u}nchen, Germany.
    This work has been supported by the Alexander von Humboldt Foundation with funds from the German Federal Ministry of Education and Research (BMBF).  E-mail addresses: {\tt \{marinus.gottschau,felix.happach,marcus.kaiser,clara.waldmann\}@tum.de}
    }
     \and Felix Happach\footnotemark[\value{footnote}]
     \and Marcus Kaiser\footnotemark[\value{footnote}]
     \and Clara Waldmann\footnotemark[\value{footnote}]
     }
\date{}

\begin{document}

    \maketitle


\abstract{\minbudget~is a scheduling problem with precedence constraints, i.e., a scheduling problem on a partially ordered set of jobs $(N, \unlhd)$.
A job $j \in N$ is available for scheduling, if all jobs $i \in N$ with $i \unlhd j$ are completed.
Further, each job $j \in N$ is assigned real valued costs $c_{j}$, which can be negative or positive.
A schedule is an ordering $j_1, \dots, j_{\vert N \vert}$ of all jobs in $N$. The budget of a schedule is the external investment needed to complete all jobs, i.e., it is $\max_{l \in \{0, \dots, \vert N \vert \} } \sum_{1 \le k \le l} c_{j_k}$. The goal is to find a schedule with minimum budget.
Rafiey et al.~(2015) showed that \minbudget~is \NPhard~following from a reduction from a molecular folding problem. We extend this result and prove that it is \NPhard~to $\alpha(N)$-approximate the minimum budget even on bipartite partial orders.
We present structural insights that lead to arguably simpler algorithms and extensions of the results by Rafiey et al.~(2015).
In particular, we show that there always exists an optimal solution that partitions the set of jobs and schedules each subset independently of the other jobs.
We use this structural insight to derive polynomial-time algorithms that solve the problem to optimality on series-parallel and convex bipartite partial orders.

\paragraph{Note:}
\textit{This work was presented at MAPSP 2019.
The authors thank Thomas Lidbetter for bringing previous results to their attention.
The problem at hand is known as minimizing maximum cumulative cost.
Abdel-Wahab and Kameda \cite{AbdelKemeda1978} and Monma and Sidney \cite{MonmaSidney1979} examine the problem for series-parallel precedence constraints.
Decompositions very similar to the one developed in this paper are deduced by Abdel-Wahab and Kameda \cite{Abdel-Wahab1980} and Sidney \cite{Sidney1981} for a more general class of objective functions.
}

\section{Introduction}
The following scheduling problem arises in the design of programmed nucleic acid systems \cite{Manuch2011}.
We are given a finite set $N$ of jobs with $\abs{N} = n$.
Each job~$j$ has some costs $c_j$, which can be positive or negative.
A positive value may be interpreted as an investment that has to be made to execute job~$j$. A negative value may model a profit job~$j$ yields upon completion.
In the context of nucleic acid systems, jobs correspond to breaking or bonding of Watson-Crick base pairs and $c_j$ corresponds to the amount of energy needed or released, respectively.

A \emph{schedule} $S$ is a total order of the jobs in $N$. Enumerate the jobs in $N$ as $j_1, \ldots, j_n$ according to $S$.
The \emph{budget of the schedule} $S$ is defined as $b(S) = \max_{l \in \set{0, \ldots, n} } \sum_{1 \le k \le l} c_{j_k}$, which is the external investment needed to complete all jobs.
The goal is to find a schedule with minimum budget.

In constrast to other scheduling problem, see e.g.~\cite{schedulingsurvey}, we do not care about processing times of the jobs.
This problem would have a trivial solution (scheduling all jobs with negative costs first and then all other jobs in any order) were it not for precedence constraints, which restrict the order in which jobs may be executed.
Precedence constraints are represented by a partial order $\trianglelefteq$ on $N$.
A job can be scheduled only after all its predecessors have been executed.
We call a schedule \emph{feasible}, if it is a linear extension of $\trianglelefteq$.
An instance of \minbudget~is given by the tuple $(N,\trianglelefteq,c)$, where $(N,\trianglelefteq)$ is a partially ordered set representing the precedence constraints on the jobs and $c \in \Rbb^N$ represents the costs of the jobs.

\paragraph{Related Work}
A special case of the problem we consider is the \textsc{Energy Barrier} problem \cite{Manuch2011,Thachuk2010}.
Ma\v{n}uch~et~al.~\cite{Manuch2011} showed that this problem is NP-complete by a reduction from \textsc{3-Partition}.
Thachuck~et~al.~\cite{Thachuk2010} present an exponential exact algorithm for the \textsc{Energy Barrier} problem.

Based on the NP-completeness proof of \cite{Manuch2011}, \minbudget~was shown to be NP-hard by Rafiey~et~al.~\cite{Rafiey2015}, even if the partial order is bipartite. They also gave polynomial-time algorithms for instances with partial order $\trianglelefteq$ that can be represented by a bipartite permutation graph, trivially perfect bipartite graph, or co-bipartite graph.

\paragraph{Our Results}
In our work, we present structural insights that lead to arguably simpler algorithms and extensions of the results by Rafiey~et~al.~\cite{Rafiey2015}.
In general, the \minbudget~problem is intrinsically inapproximable, as we will discuss in \Cref{sec:hardness}.
Hence, we focus on structural results of optimal schedules in \Cref{sec:structure} and use these to derive polynomial-time algorithms for restricted input in \Cref{sec:algorithms}. In the following, we give a short description of the respective sections.

In \Cref{sec:hardness}, we use a similar construction as \cite{Manuch2011,Rafiey2015}, to show that it is \NPhard~to decide whether the budget is zero or strictly positive for bipartite instances.
The hardness of bipartite instances is not surprising, as we show that any instance of \minbudget~can be transformed into an equivalent bipartite instance in polynomial time.
Further, we observe that the inapproximability even holds for unit costs.

In some instances, the budget of an optimal schedule might be determined by a short prefix. Then, the order of the jobs afterwards might be irrelevant and, hence, quite arbitrary. We want to utilize a subset of optimal solutions which exhibit more structure in order to devise recursive algorithms.
In \Cref{sec:structure}, we analyze the structure of an optimal solution using two characteristic quantities of schedules, we call the \emph{budget} and the \emph{return} (see \Cref{sec:budgetreturn} for a definition).
Based on these quantities, we propose a preference on the subsets of jobs by means of a preorder, the \emph{cbr-preorder}.
We show that scheduling subsets of jobs according to the preorder yields a schedule of minimum budget, and present a generic algorithm in \Cref{sec:cbrorder}.
However, feasibility of this schedule strongly depends on the choice of the subsets.
We propose a partition of the set of jobs into \emph{irreducible sets} and analyze their properties in \Cref{sec:irreducibleintervals}.
Finally in \Cref{sec:optimalityofalgorithm}, we prove that there always exists an optimal solution that schedules these irreducible sets contiguously and in increasing order w.r.t.~the aforementioned preorder.
A key result is that each irreducible set can be scheduled optimally and independently of the other jobs.
We say that these schedules are in \emph{increasing irreducible structure}, and show that any such schedule is optimal.

In \Cref{sec:algorithms}, we use our structural results to derive polynomial-time algorithms for series-parallel (\Cref{sec:seriesparallel}) and convex bipartite partial orders (\Cref{sec:convex}).
For a series-parallel partial order, we compute its decomposition tree and recursively concatenate and merge optimal schedules of the components.
We show that, if the subschedules of the components are in increasing irreducible structure, then the resulting schedule is as well.
If the partial order is convex bipartite, we propose a dynamic program.
The algorithm uses an observation that, for the correct choice of the first job in an optimal schedule, there is an optimal schedule of a series-parallel instance that coincides with the optimal schedule of the initial instance.

We conclude with some remarks in \Cref{sec:conclusion}.


\section{Hardness of Approximation}\label{sec:hardness}
As a generalization of the \NPhard~\textsc{Energy Barrier} problem considered in \cite{Manuch2011}, the \minbudget~problem is \NPhard~\cite{Rafiey2015}.
Using a similar construction, we can show that \minbudget~cannot be approximated within a factor $\alpha(n)$, where $\alpha(n)$ does only depend on the number of jobs $n$ or the precedence constraints, but not on the costs.

	%

	\begin{lemma}[Hardness of Approximation]
	\label{lem:andbudgetminimization:hardness}
	Let $(N,\unlhd,c)$ be an instance of \minbudget~where $\unlhd$ is bipartite.
	For any $\alpha = \alpha(N) \ge 1$, it is \NPhard~to $\alpha$-approximate the minimum budget.

	\begin{proof}
	We will reduce \textsc{Energy Barrier} to \minbudget.
	An instance $(\Ical,\Fcal,w)$ of the \textsc{Energy Barrier} problem is given by two laminar systems $\Ical, \Fcal$ of closed intervals in $\Rbb$ and weights $w \in \Rbbp^{\Ical \cup \Fcal}$.
	The task is to find $k \in \Nbb_0$ and a sequence $(\Ical = \Ccal_0, \Ccal_1, \ldots, \Ccal_k = \Fcal)$ of laminar systems $\Ccal_i \subseteq \Ical \cup \Fcal$ such that $\card{\Ccal_{i-1} \triangle \Ccal_{i}} = 1$ for all $i \in [k]$ that minimizes the energy barrier $\max_{i \in \set{0, \ldots, k}} \braces{w(\Ccal_0) - w(\Ccal_i)}$, where $w(\Ccal) = \sum_{j \in \Ccal} w_j$.
	Here $\Ccal_{i-1} \triangle \Ccal_{i} := \braces{\Ccal_{i-1} \setminus \Ccal_{i} } \cup \braces{\Ccal_{i} \setminus \Ccal_{i-1}}$ denotes the symmetric difference of the two sets.
		By scaling appropriately, we may assume that the weights $w$ are integer.
		Further, let $B \in \Nbb$.
		We will construct an instance of \minbudget~which has budget zero if and only if $(\Ical,\Fcal,w)$ admits an energy barrier no larger than~$B$.


		Define the set of jobs $N := (\Ical \triangle \Fcal) \cupdot \set{j_{B}}$ and the partial order $\unlhd$ on $N$ by $I \unlhd F$ if and only if  $I \in \Ical \setminus \Fcal$, $F \in \Fcal \setminus \Ical$, $I \setminus F \ne \emptyset$, $F \setminus I \ne \emptyset$, and $F \cap I \ne \emptyset$.
		Set the costs $c \in \Rbb^N$ by $c_I := w_I$ for all $I \in \Ical \setminus \Fcal$, $c_F := -w_F$ for all $F \in \Fcal \setminus \Ical$, and $c_{j_B} := -B$.

		Assume there is $k \in \Nbb_0$ and a sequence $(\Ical = \Ccal_0, \Ccal_1, \ldots, \Ccal_k = \Fcal)$ of laminar systems $\Ccal_i \subseteq \Ical \cup \Fcal$ such that $\card{\Ccal_{i-1} \triangle \Ccal_{i}} = 1$ for all $i \in [k]$ and the energy barrier $\max_{i \in \set{0, \ldots, k}} (w(\Ccal_0) - w(\Ccal_i))$ is no greater than $B$.
		Choose the one that minimizes $k$.
		Then the singletons $\Ccal_{i-1} \triangle \Ccal_i$ for $i \in [k]$ form a partition of $\Ical \triangle \Fcal$. Enumerate the corresponding jobs in $N$ such that $\{j_i\} = \Ccal_{i-1} \triangle \Ccal_i$.
		Then scheduling the jobs in order $j_B, j_1, \dots, j_{k}$ is feasible and yields a budget of zero.

		Consider a feasible schedule with budget strictly less than one.
		Since the budget is determined by a sum of costs and all costs are integer, the budget is equal to zero.
		Let $k := \card{\Ical \triangle \Fcal}$ and $\Ical \triangle \Fcal = \set{j_1, \ldots, j_k}$ be indexed such that the jobs appear in order $j_1,  j_2, \dots, j_k$ in the schedule.
		Due to the construction of $\unlhd$, setting $\Ccal_i := \Ical \triangle \set{j_1, \ldots, j_i}$ for all $i \in \set{0, \ldots, k}$ yields a sequence of laminar systems.
		Noticing that
		\begin{equation*}
		w(\Ccal_0) - w (\Ccal_i) = w(\Ical) - \braces{w(\Ical) - w \braces{\Ical \cap \set{j_1, \ldots, j_i}} + w \braces{\Fcal \cap \set{j_1, \ldots, j_i}} } = \sum_{l = 1}^i c_{j_l}
		\end{equation*}
		is true for every $i \in \set{0, \ldots, k}$ yields an energy barrier no greater than $B$.

		Hence, distinguishing between a budget of zero or a strictly positive budget is computationally equivalent to solving the \textsc{Energy Barrier} problem.
	\end{proof}
	\end{lemma}

Note that the instance we construct in the proof is an instance of the \textsc{Direct Set Barrier Problem} of \cite{Thachuk2010}.
The statement of \cref{lem:andbudgetminimization:hardness} remains true, if we restrict the costs to unit costs $c \in \{-1,+1\}^N$.
In that case a trivial upper bound on the budget is the number of jobs $n$.
Hence instead of adding a single job with cost $-B$, one could add $B \leq n$ jobs with cost $-1$.

The fact that \minbudget~is already \NPhard~for bipartite partial orders is not surprising.
We can transform any instance to a bipartite instance with the same minimum budget in polynomial time.

Recall that a feasible schedule $S$ is a linear extension of the partial order $\unlhd$.
We write $i \mathbin{S} j$ if $i$ is scheduled before $j$ in the schedule $S$, i.e.,\ $i$ is less than $j$ w.r.t.~$S$.

\begin{theorem}[Reduction to Bipartite Partial Orders]
    \label{theorem:bipartite:reduction}
        Let $(N, \unlhd,c)$ be an instance of \minbudget.
        Let $N^+ := \set{j \in N: c_j \ge 0}$ and $N^- := \set{j \in N: c_j < 0}$ and set $\unlhd_1 := \unlhd \cap (N^+ \times N^-)$.

        Then the minimal budget of $(N, \unlhd, c)$ is the same as the minimal budget of $(N^+ \cupdot N^-, \unlhd_1, c)$.
        \begin{proof}
            Note that any feasible schedule for $(N, \unlhd)$ is also feasible for $(N^+ \cupdot N^-, \unlhd_1)$.
            So the bipartite instance $(N^+ \cupdot N^-, \unlhd_1, c)$ is a relaxation of the initial instance.

            In order to show equivalence, we show that any feasible schedule for $(N^+ \cupdot N^-, \unlhd_1)$ can be modified to obey the partial order $\unlhd$ without increasing the budget.
            For a schedule $S$ of $N$, define the potential $\varphi(S) := \card{ \set{ (i, j) \in N \times N: i \mathbin{S} j \land j \unlhd i } }$ to be the number of pairs in $S$ violating the partial order $\unlhd$.
            Let $S_1$ be a feasible schedule of $(N^+ \cupdot N^-, \unlhd_1)$ minimizing~$\varphi(S_1)$ among all feasible schedules with minimal budget.
            That is for all feasible schedules $S$ of $(N^+ \cupdot N^-, \unlhd_1)$ with $b(S) = b(S_1)$, it holds $\varphi(S_1) \le \varphi(S)$.
            We claim that $\varphi(S_1) = 0$, i.e.~$S_1$ is a feasible schedule for $(N, \unlhd)$ with budget at most the minimal budget of the bipartite instance $(N^+ \cupdot N^-, \unlhd_1)$.

            Assume for a contradiction that $\varphi(S_1) > 0$.
            Consider $k, l \in N$ with $k \mathbin{S_1} l$ and $l \unlhd k$ such that $\card{ \set{ j \in N: k \mathbin{S_1} j \mathbin{S_1} l } }$ is minimal.
            In other words, $k$ and $l$ are two closest jobs in the schedule $S_1$ that are in wrong order with respect to $\unlhd$.
            Since $S_1$ is a feasible schedule of $(N^+ \cupdot N^-, \unlhd_1)$, it has to hold that $k \in N^+$ or $l \in N^-$.
            \\
            Assume $l \in N^-$ is true.
            We define another feasible schedule $S_2$ for $(N^+ \cupdot N^-, \unlhd_1)$ that has the same budget as $S_1$ but strictly less potential. Thus, contradicting the choice of $S_1$.

            Define schedule $S_2$ based on $S_1$ by making $l$ immediate predecessor of $k$ while preserving the order of $N \setminus \set{l}$.
            Hereby, the order of $l$ and $k$ is switched.
            See \cref{fig:lbeforek} for an illustration of the two schedules $S_1$ and $S_2$.
           Formally, we define $S_2$ by:
            \begin{align*}
                S_2 := &\set{(j,l) : j \mathbin{S_1} k} \cupdot
                       \set{(l,j): k \mathbin{S_1} j \land j \ne l} \cupdot \set{(l,k)} \cupdot \\
                       &\set{(j,k): j \mathbin{S_1} k} \cupdot
                       \set{(k,j): k \mathbin{S_1} j \land j \ne l} \cupdot \\
                       &\set{(i,j): i \mathbin{S_1} j \land i,j \notin \set{k,l}}
            \end{align*}

            \begin{figure}[h]
                \centering
                \begin{tikzpicture}[ipe stylesheet]
  \draw
    (192, 776) circle[radius=8];
  \draw
    (288, 776) circle[radius=8];
  \draw
    (192, 752) circle[radius=8];
  \draw
    (216, 752) circle[radius=8];
  \node[ipe node]
     at (189.631, 773.044) {$k$};
  \node[ipe node]
     at (190.355, 749.044) {$l$};
  \node[ipe node]
     at (213.726, 748.921) {$k$};
  \node[ipe node]
     at (286.331, 773.044) {$l$};
  \filldraw[rgb color={fill=0.753 0.753 0.753}]
    (208, 784) rectangle (272, 768);
  \filldraw[rgb color={fill=0.753 0.753 0.753}]
    (232, 760) rectangle (296, 744);
  \node[ipe node]
     at (211.338, 773.797) {$\{k \mathbin{S_1} j \mathbin{S_1} l \}$};
  \filldraw[fill=white]
    (304, 784) rectangle (352, 768);
  \filldraw[fill=white]
    (128, 784) rectangle (176, 768);
  \node[ipe node]
     at (133.338, 773.797) {$\{j \mathbin{S_1} k \}$};
  \node[ipe node]
     at (309.338, 773.797) {$\{l \mathbin{S_1} j\}$};
  \filldraw[fill=white]
    (304, 760) rectangle (352, 744);
  \filldraw[fill=white]
    (128, 760) rectangle (176, 744);
  \draw[-{>[ipe arrow small]}]
    (288, 784)
     .. controls (253.3333, 805.3333) and (218.6667, 805.3333) .. (184, 784);
\end{tikzpicture}
                \caption{The schedules $S_1$ (top) and $S_2$ (bottom), if $l \in N^-$.}
                \label{fig:lbeforek}
            \end{figure}
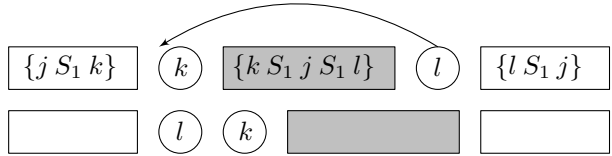

            Note that there is no $j \in N$ such that $k \mathbin{S_1} j \mathbin{S_1} l$ and $j \unlhd l \unlhd k$, as this would contradict the choice of $k$ and $l$.
            This yields that $S_2$ is indeed feasible for~$(N^+ \cupdot N^-, \unlhd_1)$.
            Further, it follows that $\varphi(S_2) = \varphi(S_1) - 1$.

            For $j \in N$, it holds:
            \begin{equation*}
                c_j + \sum_{i \in N: i \mathbin{S_2} j} c_i \le
                \begin{cases}
                    c_j + \sum_{i \in N: i \mathbin{S_1} j} c_i & \text{for }j \ne l\\
                    c_j + \sum_{i \in N: i \mathbin{S_1} k} c_i & \text{for }j = l
                \end{cases}
            \end{equation*}
            and, hence, $b(S_2) \le b(S_1)$.

            In total, $S_2$ is a feasible schedule for $(N^+ \cupdot N^-, \unlhd_1)$ with $b(S_2) = b(S_1)$ (as $S_1$ has minimal budget) and $\varphi(S_2) < \varphi(S_1)$.
            This contradicts the choice of $S_1$.
            \\
            The case $k,l \in N^+$ can be handled in a similar way by making $k$ immediate successor of $l$.
            See \cref{fig:kafterl} for an illustration of this case.
            \begin{figure}[h]
                \centering
                \begin{tikzpicture}[ipe stylesheet]
  \draw
    (192, 776) circle[radius=8];
  \draw
    (288, 776) circle[radius=8];
  \draw
    (264, 752) circle[radius=8];
  \draw
    (288, 752) circle[radius=8];
  \node[ipe node]
     at (189.631, 773.044) {$k$};
  \node[ipe node]
     at (262.355, 749.044) {$l$};
  \node[ipe node]
     at (285.726, 748.921) {$k$};
  \node[ipe node]
     at (286.331, 773.044) {$l$};
  \filldraw[rgb color={fill=0.753 0.753 0.753}]
    (208, 784) rectangle (272, 768);
  \filldraw[rgb color={fill=0.753 0.753 0.753}]
    (184, 760) rectangle (248, 744);
  \node[ipe node]
     at (211.338, 773.797) {$\{k \mathbin{S_1} j \mathbin{S_1} l \}$};
  \filldraw[fill=white]
    (304, 784) rectangle (352, 768);
  \filldraw[fill=white]
    (128, 784) rectangle (176, 768);
  \node[ipe node]
     at (133.338, 773.797) {$\{j \mathbin{S_1} k \}$};
  \node[ipe node]
     at (309.338, 773.797) {$\{l \mathbin{S_1} j\}$};
  \filldraw[fill=white]
    (304, 760) rectangle (352, 744);
  \filldraw[fill=white]
    (128, 760) rectangle (176, 744);
  \draw[-{>[ipe arrow small]}]
    (192, 784)
     .. controls (226.6667, 805.3333) and (261.3333, 805.3333) .. (296, 784);
\end{tikzpicture}
                \caption{The schedules $S_1$ (top) and $S_2$ (bottom), if $k, l \in N^+$.}
                \label{fig:kafterl}
            \end{figure}
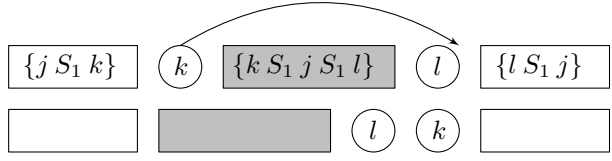
        \end{proof}
    \end{theorem}

    \begin{corollary}[Polynomial Equivalence]
    \label{theorem:bipartite:equivalence}
        The \minbudget~problem and its variant when restricting to bipartite partial orders are polynomially equivalent.
        \begin{proof}
            To solve the instance $(N, \unlhd, c)$ we construct the bipartite instance $(N^+ \cupdot N^-, \unlhd_1, c)$ as in \Cref{theorem:bipartite:reduction}.
            Given a solution to $(N^+ \cupdot N^-, \unlhd_1, c)$, the proof of \cref{theorem:bipartite:reduction} gives a constructive method for minimizing the considered potential function while not increasing the budget.
            The resulting schedule is feasible for the original instance.
            The involved operations can be performed in polynomial time, since the used potential function is integer, non-negative, and bounded by the size of the input.
        \end{proof}
    \end{corollary}

Despite these hardness results, we look at the structure of some optimal solutions in the following section.
This structure will help to design algorithms for instances with special classes of partial orders (\Cref{sec:algorithms}).


\section{Structure of Optimal Schedules}\label{sec:structure}


In this section, we introduce notations that are used in the rest of this paper and define the main building blocks of our structural results and the algorithms for the special cases.

Our idea to solve \minbudget~is to partition the jobs into blocks which we would like to schedule contiguously.
Every block is scheduled optimally without taking into account the other blocks.
To obtain a schedule of all jobs, we have to find an ordering of the optimal schedules of these blocks.
For this, we define the \emph{cbr-preorder} on subsets of jobs in \Cref{sec:cbrorder} and show that there is an optimal schedule that schedules the subsets of jobs in increasing order w.r.t.~the cbr-preorder.
For a fixed partition of the jobs, the cbr-preorder determines the best schedule that schedules all part contiguously.
To obtain an optimal feasible schedule for all jobs, we choose a special partition of the jobs into \emph{irreducible intervals}, see \Cref{sec:irreducibleintervals}.
Finally, we show in \Cref{sec:optimalityofalgorithm} that there always exists a feasible schedule in \emph{increasing irreducible structure} and that those schedules are optimal.

We will use the standard definiton of ideals and filters of a partial order (see \cite{Mohring1989}):
\begin{definition}[Ideals, Filters, and Intervals]
    Let $\unlhd$ be a partial order on a set $N$.
    \begin{definitionenum}
        \item
        A subset $J \subseteq N$ is an \emph{ideal} of $\unlhd$ if for all $j \in J$ and all $i \in N$ with $i \unlhd j$ it holds that $i \in J$.
        \item
        A subset $F \subseteq N$ is a \emph{filter} of $\unlhd$ if for all $i \in F$ and all $j \in N$ with $i \unlhd j$ it holds that $j \in F$.
        \item
        A subset $I \subseteq N$ is an \emph{interval} of $\unlhd$ if there is an ideal $J \subseteq N$ and a filter $F \subseteq N$ of $\unlhd$ such that $I = J \cap F$.
    \end{definitionenum}
\end{definition}


\subsection{Budget and Return}\label{sec:budgetreturn}

For schedules, we use the following notation.
We call an ideal of a schedule $S$ a \emph{prefix} of $S$, and a filter of $S$ is called a \emph{suffix} of $S$.
Thinking of a schedule as a list of jobs, a prefix is an initial segment of the schedule and a suffix is a final segment.
The \emph{concatenation} of two schedules $S_1$ and $S_2$ on disjoint sets of jobs is denoted by $S_1 \oplus S_2$.
The \emph{cost} of a subset of jobs $I \subseteq N$ is defined as $c(I): = \sum_{j \in I} c_j$, where $c(\emptyset) = 0$.
If $S$ is a schedule of jobs $I \subseteq N$, we denote its cost by $c(S) := c(I)$.
Note that the cost of a schedule is independent of the ordering of the jobs in $I$.
Sometimes we will use the informal notation $S \cap I$ to denote the subschedule of $S$ that contains only jobs of $I \subseteq N$ (in the same order as $S$).
Similarly, $S \setminus I$ is used to refer to the subschedule of $S$ that contains only jobs of $N \setminus I$ (in the same order as $S$).
We start out by defining two characteristic quantities of schedules, the budget and the return.

\begin{definition}[Budget and Return of Schedules]
\label{def:budgetreturnorder}
    Let $S$ be a schedule of jobs $N$.
    \begin{definitionenum}
        \item \label{def:budgetreturnorder:budget}
        The \emph{budget $b(S)$} of schedule $S$ is defined as the maximal cost of a prefix of $S$, i.e.
        \begin{equation*}
            b(S) := \max \set{ c(P) : P \text{ is prefix of } S } .
        \end{equation*}
        \item \label{def:budgetreturnorder:return}
        The \emph{return $r(S)$} of schedule $S$ is defined as $r(S) := c(S) - b(S).$
    \end{definitionenum}
\end{definition}

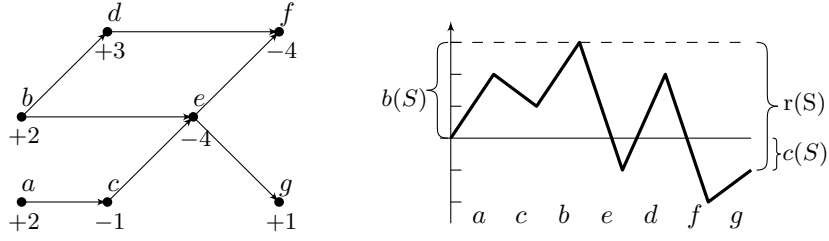
\begin{figure}
	\centering
	\begin{tikzpicture}[ipe stylesheet] 
  \pic
     at (80, 752) {ipe disk};
  \pic
     at (80, 720) {ipe disk};
  \pic
     at (112, 720) {ipe disk};
  \pic
     at (112, 784) {ipe disk};
  \pic
     at (144, 752) {ipe disk};
  \pic
     at (176, 720) {ipe disk};
  \pic
     at (176, 784) {ipe disk};
  \draw[-{>[ipe arrow small]}]
    (80, 720)
     -- (112, 720);
  \draw[-{>[ipe arrow small]}]
    (112, 720)
     -- (144, 752);
  \draw[-{>[ipe arrow small]}]
    (144, 752)
     -- (176, 720);
  \draw[-{>[ipe arrow small]}]
    (80, 752)
     -- (144, 752);
  \draw[-{>[ipe arrow small]}]
    (112, 784)
     -- (176, 784);
  \node[ipe node]
     at (112, 788) {$d$};
  \node[ipe node]
     at (176, 788) {$f$};
  \node[ipe node]
     at (144, 756) {$e$};
  \node[ipe node]
     at (176, 724) {$g$};
  \node[ipe node]
     at (112, 724) {$c$};
  \node[ipe node]
     at (80, 724) {$a$};
  \node[ipe node]
     at (80, 756) {$b$};
  \node[ipe node, font=\small]
     at (171, 774) {$-4$};
  \node[ipe node, font=\small]
     at (171, 710) {$+1$};
  \node[ipe node, font=\small]
     at (107, 710) {$-1$};
  \node[ipe node, font=\small]
     at (139, 741) {$-4$};
  \node[ipe node, font=\small]
     at (107, 774) {$+3$};
  \node[ipe node, font=\small]
     at (75, 742) {$+2$};
  \node[ipe node, font=\small]
     at (75, 710) {$+2$};
  \draw
    (236, 744)
     -- (352, 744);
  \node[ipe node]
     at (248, 712) {$a$};
  \node[ipe node]
     at (264, 712) {$c$};
  \node[ipe node]
     at (280, 712) {$b$};
  \node[ipe node]
     at (296, 712) {$e$};
  \node[ipe node]
     at (312, 712) {$d$};
  \node[ipe node]
     at (328, 712) {$f$};
  \node[ipe node]
     at (344, 712) {$g$};
  \draw
    (240, 756)
     -- (244, 756);
  \draw
    (240, 768)
     -- (244, 768);
  \draw[ipe pen fat]
    (240, 744)
     -- (256, 768)
     -- (272, 756)
     -- (288, 780)
     -- (304, 732)
     -- (320, 768)
     -- (336, 720)
     -- (352, 732);
  \draw
    (240, 732)
     -- (244, 732);
  \draw
    (240, 720)
     -- (244, 720);
  \draw[-{>[ipe arrow small]}]
    (240, 712)
     -- (240, 788);
  \draw
    (240, 780)
     -- (244, 780);
  \node[ipe node, font=\small]
     at (214.147, 755.986) {$b(S)$};
  \draw[ipe dash dashed]
    (240, 780)
     -- (352, 780);
  \node[ipe node, font=\small]
     at (363.637, 735.855) {$c(S)$};
  \draw
    (354, 732)
     .. controls (358, 732) and (358, 734) .. (358, 737.6667)
     .. controls (358, 741.3333) and (358, 746.6667) .. (358, 750.3333)
     .. controls (358, 754) and (358, 756) .. (362, 756);
  \draw
    (362, 756)
     .. controls (358, 756) and (358, 758) .. (358, 761.6667)
     .. controls (358, 765.3333) and (358, 770.6667) .. (358, 774.3333)
     .. controls (358, 778) and (358, 780) .. (354, 780);
  \node[ipe node, font=\small]
     at (364.073, 754.477) {$$r(S)$$};
  \draw[-{>[ipe arrow small]}]
    (80, 752)
     -- (112, 784);
  \draw[-{>[ipe arrow small]}]
    (144, 752)
     -- (176, 784);
  \draw[shift={(360, 744)}, scale=0.375]
    (0, 0)
     .. controls (4, 0) and (4, -2) .. (4, -4.3333)
     .. controls (4, -6.6667) and (4, -9.3333) .. (4, -11.6667)
     .. controls (4, -14) and (4, -16) .. (8, -16);
  \draw[shift={(363, 738)}, scale=0.375]
    (0, 0)
     .. controls (-4, 0) and (-4, -2) .. (-4, -4.3333)
     .. controls (-4, -6.6667) and (-4, -9.3333) .. (-4, -11.6667)
     .. controls (-4, -14) and (-4, -16) .. (-8, -16);
  \draw
    (239, 780)
     .. controls (235, 780) and (235, 778) .. (235, 775.3333)
     .. controls (235, 772.6667) and (235, 769.3333) .. (235, 766.6667)
     .. controls (235, 764) and (235, 762) .. (231, 762);
  \draw
    (231, 762)
     .. controls (235, 762) and (235, 760) .. (235, 757.3333)
     .. controls (235, 754.6667) and (235, 751.3333) .. (235, 748.6667)
     .. controls (235, 746) and (235, 744) .. (239, 744);
\end{tikzpicture}
  \vspace{-0.5em}
	\caption{Instance of \minbudget~given by the transitively reduced graph of $\unlhd$ (left) and a feasible schedule $S$ (right). The horizontal axis indicates the order in which jobs are scheduled, the vertical axis represents the total cost of the prefix. The cost, budget, and return of $S$ evaluate to $c(S)=-1, b(S)=3, r(S)=-4 $.
  }
    \label{fig:budgetreturn}
\end{figure}

\cref{fig:budgetreturn} illustrates the cost, budget, and return of a schedule.
Note that the budget and return of a schedule are by definition non-negative and non-positive, respectively.
The return of $S$ can equivalently be expressed as $r(S) = \min \set{ c(Q) : Q \text{ is suffix of } S }$.
From the definitions, we get the following observation on the budget and the return of concatenated schedules on disjoint job sets.

\begin{observation}[Budget and Return of Concatenations]
\label{obs:budgetreturnconcat}
  Let $(N, \unlhd, c)$ be an instance of \minbudget~and let $N_1, N_2 \subseteq N$ with $N_1 \cap N_2 = \emptyset$.
  Let $S_1$ and $S_2$ be schedules of the jobs in $N_1$ and $N_2$, respectively.

  Then the budget and return of $S_1 \oplus S_2$ satisfy
  \begin{observationenum}
      \item \label{obs:budgetreturncost:composition:budget}
      $b(S_1 \oplus S_2) = \max \set{b(S_1), c(S_1) + b(S_2)}$ and
      \item \label{obs:budgetreturncost:composition:return}
      $r(S_1 \oplus S_2) = \min \set{r(S_1) + c(S_2), r(S_2)}$.
\end{observationenum}
\end{observation}

We denote by $(I,\unlhd_I,c)$ the restriction of the instance $(N,\unlhd,c)$ to a set of jobs $I \subseteq N$.

To compare subsets of jobs, we lift the definitions of budget and return to such subinstances by optimizing over all feasible schedules.

\begin{definition}[Budget and Return for Sets of Jobs]
\label{def:budgetreturninterval}
    Let $(N,\unlhd,c)$ be an instance of \minbudget . For a subset $I \subseteq N$, we define
    \begin{definitionenum}
        \item \label{def:preorder:graphs:budget}
        the \emph{budget} of $I$ as $b(I) := \min \set{b(S) : S \text{ feasible schedule of } (I, \unlhd_I, c)}$,
        \item \label{def:preorder:graphs:return}
        the \emph{return} of $I$ as $r(I) := c(I) - b(I)$.
    \end{definitionenum}
\end{definition}

Note that $c(\emptyset) = b(\emptyset) = r(\emptyset) = 0$. We can again express the return equivalently as $r(I) = \max\{r(S) : S \text{ feasible schedule of } (I, \unlhd_I, c)\}$.
We call a feasible schedule $S$ of $(I, \unlhd_I, c)$ an \emph{optimal} schedule of $(I, \unlhd_I,c)$ (or just $I$), if $S$ solves the optimization problem that defines $b(I)$, i.e., $b(S) = b(I)$.
Often we denote an optimal schedule of $I$ by $S^*_I$.
In these terms, the goal of \minbudget~is to determine $b(N)$ and $S^*_N$.

\subsection{The cbr-Preorder on Subsets of Jobs}\label{sec:cbrorder}

We now define a preorder on subsets of jobs which gives us the best ordering of the schedules of the blocks for a fixed partition of the jobs.
However, this schedule strongly depends on the partition, so for a different partition of the jobs there may be better overall schedules of $N$.
How to choose the partition (the blocks) in an optimal and feasible way is discussed in \Cref{sec:irreducibleintervals}.

The idea is to decide on an ordering of the optimal schedules of $I \subseteq N$ and $I^\prime \subseteq N$ in an overall schedule that minimizes the total budget. We propose a preference on subsets of jobs by comparing their respective costs, budget, and return.

\begin{definition}[cbr-Preorder for Sets of Jobs]
\label{def:cbrpreorder}
    Let $(N,\unlhd,c)$ be an instance of \minbudget.
        We define the \emph{cbr-preorder} $\preceq$ on subsets of~$N$ as follows.
        For $I, I^\prime \subseteq N$, we define $I \preceq I^\prime$ if and only if
        \begin{lemmaenum}
           \item  $c(I) < 0 \le c(I^\prime)$, or
           \item  $c(I), c(I^\prime) < 0$ and $b(I) < b(I^\prime)$, or
           \item  $c(I), c(I^\prime) < 0$ and $b(I) = b(I^\prime)$ and $r(I) \leq r(I^\prime)$, or
           \item  $c(I), c(I^\prime) \ge 0$ and $r(I) \leq r(I^\prime)$.
        \end{lemmaenum}
\end{definition}

Recall that the empty set satisfies $c(\emptyset) = b(\emptyset) = r(\emptyset) = 0$. Hence, $I \preceq \emptyset$ for any $I \subseteq N$.
We give a brief intuition to the four cases in the above definition, see \cref{fig:cbr-preorder}. Take two disjoint subsets of jobs $I$ and $I'$.
As we want to minimze the total budget, we want to schedule job sets with negative costs as early as possible as the reward will help to decrease the budget needed for the following jobs.
If only one of $I$ and $I^\prime$ has negative cost, we want to schedule those jobs earlier.
If both job sets have negative cost, we prefer the one with smaller budget, as the reward gained by scheduling those first may decrease the budget of the other set enough to keep the total budget low.
If both sets have negative cost and need the same budget, we prefer to schedule the set with smaller return first. Recall that the return is always non-positive and, thus, we prefer the jobs with higher reward.
Finally, if both job sets have positive cost, we do not consider the budgets of the sets, but we again prefer the set with smaller return. Intuitively, the set with smaller return lowers the current costs as far as possible.

\begin{figure}
	\centering
	\resizebox{0.7\textwidth}{!}{
	\begin{tikzpicture}[ipe stylesheet]
	  \draw[shift={(221, 688)}, xscale=0.3791, yscale=0.75, ->]
	(0, 0)
	-- (364, 0);
	\draw[shift={(224, 676)}, scale=0.75, ->]
	(0, 0)
	-- (0, 76);
	\draw[shift={(77, 632)}, xscale=0.3791, yscale=0.75, ->]
	(0, 0)
	-- (364, 0);
	\draw[shift={(221, 632)}, xscale=0.3791, yscale=0.75, ->]
	(0, 0)
	-- (364, 0);
	\draw[shift={(224, 544)}, scale=0.75, ->]
	(0, 0)
	-- (0, 76);
	\draw[shift={(221, 568)}, xscale=0.3791, yscale=0.75, ->]
	(0, 0)
	-- (364, 0);
	\draw[shift={(80, 544)}, scale=0.75, ->]
	(0, 0)
	-- (0, 76);
	\draw[shift={(77, 568)}, xscale=0.3791, yscale=0.75, ->]
	(0, 0)
	-- (364, 0);
	\draw[shift={(77, 488)}, xscale=0.3791, yscale=0.75, ->]
	(0, 0)
	-- (364, 0);
	\draw[shift={(221, 488)}, xscale=0.3791, yscale=0.75, ->]
	(0, 0)
	-- (364, 0);
	\draw[shift={(224, 484)}, scale=0.75, ->]
	(0, 0)
	-- (0, 76);
	\draw[shift={(224, 608)}, scale=0.75, ->]
	(0, 0)
	-- (0, 76);
  \draw[shift={(80, 688)}, scale=0.75, rgb color={draw=1 0.502 0}, ipe pen fat]
    (0, 0)
     -- (32, 32)
     -- (48, 16);
  \draw[shift={(176, 715)}, scale=0.75, ipe dash dotted]
    (0, 0) rectangle (-80, -24);
  \draw[shift={(116, 700)}, scale=0.75]
    (0, 0)
     -- (8, 8)
     -- (16, -4)
     -- (24, 16)
     -- (32, 4);
  \draw[shift={(149, 706)}, scale=0.75]
    (0, 0)
     -- (4, 12)
     -- (16, -4)
     -- (24, 0)
     -- (28, -4)
     -- (36, 8);
  \draw[shift={(140, 703)}, scale=0.75, ipe dash dotted]
    (0, 0)
     -- (12, 4);
  \pic[fill=white]
     at (116, 700) {ipe fdisk};
  \draw[shift={(176, 712)}, scale=0.75, rgb color={draw=0 0.502 1}, ipe pen fat]
    (0, 0)
     -- (16, 16)
     -- (32, -16);
  \pic[fill=white]
     at (176, 712) {ipe fdisk};
  \draw[shift={(308, 691)}, scale=0.75, ipe dash dotted]
    (0, 0) rectangle (-80, -24);
  \draw[shift={(248, 676)}, scale=0.75]
    (0, 0)
     -- (8, 8)
     -- (16, -4)
     -- (24, 16)
     -- (32, 4);
  \draw[shift={(281, 682)}, scale=0.75]
    (0, 0)
     -- (4, 12)
     -- (16, -4)
     -- (24, 0)
     -- (28, -4)
     -- (36, 8);
  \draw[shift={(272, 679)}, scale=0.75, ipe dash dotted]
    (0, 0)
     -- (12, 4);
  \draw[shift={(308, 688)}, scale=0.75, rgb color={draw=1 0.502 0.251}, ipe pen fat]
    (0, 0)
     -- (32, 32)
     -- (48, 16);
  \draw[shift={(224, 688)}, scale=0.75, rgb color={draw=0 0.502 1}, ipe pen fat]
    (0, 0)
     -- (16, 16)
     -- (32, -16);
  \draw[shift={(78, 724)}, scale=0.75, ipe dash dashed]
    (0, 0)
     -- (364, 0);
  \draw[shift={(77, 688)}, xscale=0.3791, yscale=0.75, ->]
    (0, 0)
     -- (364, 0);
  \pic[fill=white]
     at (308, 688) {ipe fdisk};
  \pic[fill=white]
     at (248, 676) {ipe fdisk};
  \draw[shift={(176, 635)}, scale=0.75, ipe dash dotted]
    (0, 0) rectangle (-80, -24);
  \draw[shift={(116, 620)}, scale=0.75]
    (0, 0)
     -- (8, 8)
     -- (16, -4)
     -- (24, 16)
     -- (32, 4);
  \draw[shift={(149, 626)}, scale=0.75]
    (0, 0)
     -- (4, 12)
     -- (16, -4)
     -- (24, 0)
     -- (28, -4)
     -- (36, 8);
  \draw[shift={(140, 623)}, scale=0.75, ipe dash dotted]
    (0, 0)
     -- (12, 4);
  \draw[shift={(308, 623)}, scale=0.75, ipe dash dotted]
    (0, 0) rectangle (-80, -24);
  \draw[shift={(248, 608)}, scale=0.75]
    (0, 0)
     -- (8, 8)
     -- (16, -4)
     -- (24, 16)
     -- (32, 4);
  \draw[shift={(281, 614)}, scale=0.75]
    (0, 0)
     -- (4, 12)
     -- (16, -4)
     -- (24, 0)
     -- (28, -4)
     -- (36, 8);
  \draw[shift={(272, 611)}, scale=0.75, ipe dash dotted]
    (0, 0)
     -- (12, 4);
  \draw[shift={(78, 656)}, scale=0.75, ipe dash dashed]
    (0, 0)
     -- (364, 0);
  \draw[shift={(176, 577)}, scale=0.75, ipe dash dotted]
    (0, 0) rectangle (-80, -24);
  \draw[shift={(116, 562)}, scale=0.75]
    (0, 0)
     -- (8, 8)
     -- (16, -4)
     -- (24, 16)
     -- (32, 4);
  \draw[shift={(149, 568)}, scale=0.75]
    (0, 0)
     -- (4, 12)
     -- (16, -4)
     -- (24, 0)
     -- (28, -4)
     -- (36, 8);
  \draw[shift={(140, 565)}, scale=0.75, ipe dash dotted]
    (0, 0)
     -- (12, 4);
  \draw[shift={(308, 565)}, scale=0.75, ipe dash dotted]
    (0, 0) rectangle (-80, -24);
  \draw[shift={(248, 550)}, scale=0.75]
    (0, 0)
     -- (8, 8)
     -- (16, -4)
     -- (24, 16)
     -- (32, 4);
  \draw[shift={(281, 556)}, scale=0.75]
    (0, 0)
     -- (4, 12)
     -- (16, -4)
     -- (24, 0)
     -- (28, -4)
     -- (36, 8);
  \draw[shift={(272, 553)}, scale=0.75, ipe dash dotted]
    (0, 0)
     -- (12, 4);
  \draw[shift={(78, 592)}, scale=0.75, ipe dash dashed]
    (0, 0)
     -- (364, 0);
  \draw[shift={(176, 515)}, scale=0.75, ipe dash dotted]
    (0, 0) rectangle (-80, -24);
  \draw[shift={(116, 500)}, scale=0.75]
    (0, 0)
     -- (8, 8)
     -- (16, -4)
     -- (24, 16)
     -- (32, 4);
  \draw[shift={(149, 506)}, scale=0.75]
    (0, 0)
     -- (4, 12)
     -- (16, -4)
     -- (24, 0)
     -- (28, -4)
     -- (36, 8);
  \draw[shift={(140, 503)}, scale=0.75, ipe dash dotted]
    (0, 0)
     -- (12, 4);
  \draw[shift={(308, 515)}, scale=0.75, ipe dash dotted]
    (0, 0) rectangle (-80, -24);
  \draw[shift={(248, 500)}, scale=0.75]
    (0, 0)
     -- (8, 8)
     -- (16, -4)
     -- (24, 16)
     -- (32, 4);
  \draw[shift={(281, 506)}, scale=0.75]
    (0, 0)
     -- (4, 12)
     -- (16, -4)
     -- (24, 0)
     -- (28, -4)
     -- (36, 8);
  \draw[shift={(272, 503)}, scale=0.75, ipe dash dotted]
    (0, 0)
     -- (12, 4);
  \draw[shift={(78, 536)}, scale=0.75, ipe dash dashed]
    (0, 0)
     -- (364, 0);
  \draw[shift={(80, 632)}, scale=0.75, rgb color={draw=1 0.502 0}, ipe pen fat]
    (0, 0)
     -- (32, 32)
     -- (48, -16);
  \draw[shift={(308, 620)}, scale=0.75, rgb color={draw=1 0.502 0.251}, ipe pen fat]
    (0, 0)
     -- (32, 32)
     -- (48, -16);
  \draw[shift={(176, 632)}, scale=0.75, rgb color={draw=0 0.502 1}, ipe pen fat]
    (0, 0)
     -- (16, 16)
     -- (32, -32);
  \draw[shift={(224, 632)}, scale=0.75, rgb color={draw=0 0.502 1}, ipe pen fat]
    (0, 0)
     -- (16, 16)
     -- (32, -32);
  \draw[shift={(80, 568)}, scale=0.75, rgb color={draw=1 0.502 0}, ipe pen fat]
    (0, 0)
     -- (32, 24)
     -- (48, -8);
  \draw[shift={(308, 562)}, scale=0.75, rgb color={draw=1 0.502 0.251}, ipe pen fat]
    (0, 0)
     -- (32, 24)
     -- (48, -8);
  \draw[shift={(176, 574)}, scale=0.75, rgb color={draw=0 0.502 1}, ipe pen fat]
    (0, 0)
     -- (16, 24)
     -- (32, -24);
  \draw[shift={(224, 568)}, scale=0.75, rgb color={draw=0 0.502 1}, ipe pen fat]
    (0, 0)
     -- (16, 24)
     -- (32, -24);
  \draw[shift={(80, 488)}, scale=0.75, rgb color={draw=1 0.502 0}, ipe pen fat]
    (0, 0)
     -- (32, 24)
     -- (48, 16);
  \draw[shift={(176, 512)}, scale=0.75, rgb color={draw=0 0.502 1}, ipe pen fat]
    (0, 0)
     -- (16, 32)
     -- (32, 16);
  \draw[shift={(224, 488)}, scale=0.75, rgb color={draw=0 0.502 1}, ipe pen fat]
    (0, 0)
     -- (16, 32)
     -- (32, 16);
  \draw[shift={(308, 512)}, scale=0.75, rgb color={draw=1 0.502 0.251}, ipe pen fat]
    (0, 0)
     -- (32, 24)
     -- (48, 16);
  \pic[fill=white]
     at (116, 620) {ipe fdisk};
  \pic[fill=white]
     at (176, 632) {ipe fdisk};
  \pic[fill=white]
     at (248, 608) {ipe fdisk};
  \pic[fill=white]
     at (308, 620) {ipe fdisk};
  \pic[fill=white]
     at (308, 512) {ipe fdisk};
  \pic[fill=white]
     at (308, 562) {ipe fdisk};
  \pic[fill=white]
     at (248, 550) {ipe fdisk};
  \pic[fill=white]
     at (176, 574) {ipe fdisk};
  \pic[fill=white]
     at (116, 562) {ipe fdisk};
  \pic[fill=white]
     at (116, 500) {ipe fdisk};
  \pic[fill=white]
     at (176, 512) {ipe fdisk};
  \node[ipe node]
     at (86, 703) {$I^\prime$};
  \node[ipe node]
     at (86, 644) {$I^\prime$};
  \node[ipe node]
     at (86, 577) {$I^\prime$};
  \node[ipe node]
     at (86, 500) {$I^\prime$};
  \node[ipe node]
     at (317, 524) {$I^\prime$};
  \node[ipe node]
     at (317, 574) {$I^\prime$};
  \node[ipe node]
     at (317, 635) {$I^\prime$};
  \node[ipe node]
     at (317, 703) {$I^\prime$};
  \node[ipe node]
     at (182, 537) {$I$};
  \node[ipe node]
     at (182, 592.9) {$I$};
  \node[ipe node]
     at (182, 646) {$I$};
  \node[ipe node]
     at (182, 725) {$I$};
  \node[ipe node]
     at (227, 505) {$I$};
  \node[ipe node]
     at (227, 580) {$I$};
  \node[ipe node]
     at (227, 641) {$I$};
  \node[ipe node]
     at (227, 697) {$I$};
  \node[ipe node]
     at (60, 704) {\textit{(i)}};
  \draw[shift={(80, 676)}, scale=0.75, ->]
    (0, 0)
     -- (0, 76);
  \node[ipe node]
     at (60, 636) {\textit{(ii)}};
  \draw[shift={(80, 608)}, scale=0.75, ->]
    (0, 0)
     -- (0, 76);
  \node[ipe node]
     at (60, 568) {\textit{(iii)}};
  \node[ipe node]
     at (60, 512) {\textit{(iv)}};
  \draw[shift={(80, 484)}, scale=0.75, ->]
    (0, 0)
     -- (0, 76);
  \pic[fill=white]
     at (248, 500) {ipe fdisk};
\end{tikzpicture}}
	\caption{Visualization and intuition of all four cases in \Cref{def:cbrpreorder}. After swapping $I$ and $I'$ (ignoring feasibility of the schedule) the budget does not increase. Note that the remainder of the schedule (in particular all jobs in the box) remains unchanged.}
	\label{fig:cbr-preorder}
\end{figure}
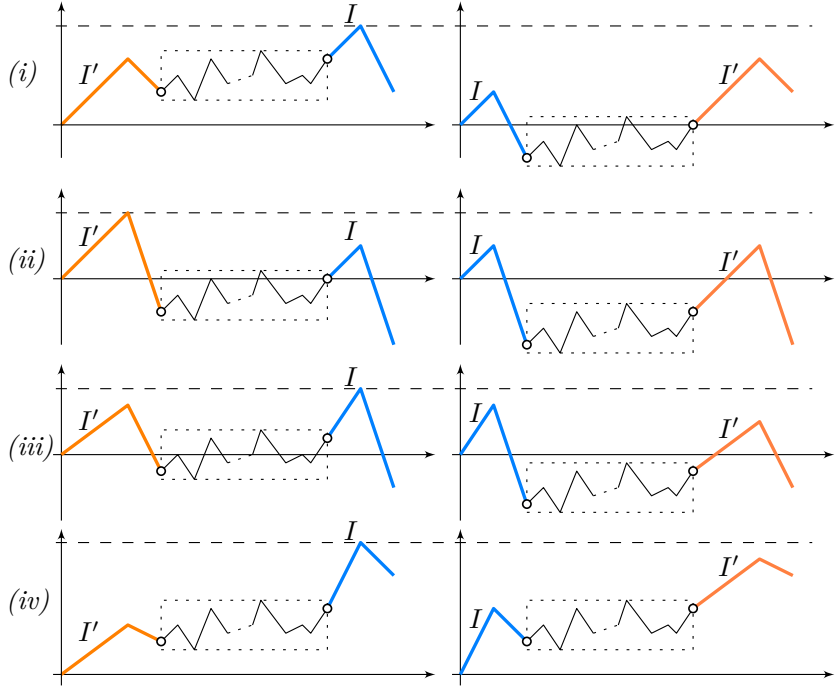

The followling lemma formalizes this intuition of the cbr-preorder.
It shows that ordering the job subsets according to the cbr-preorder is good in terms of the total budget of the schedule.
Note that the lemma does not take feasibility of schedules into account.

\begin{lemma}[Consistency of the cbr-Preorder and Budget Minimization]
\label{lem:consistency}
    Let $(N,\unlhd,c)$ be an instance of \minbudget~and $N = L \cupdot I \cupdot M \cupdot J \cupdot R $ a partition of the jobs.
    Let $S = S_L \oplus S^*_I \oplus S_M \oplus S^*_J \oplus S_R$ be a schedule of $N$. The subschedules $S_L, S_M$, and $S_R$ are arbitrary schedules of the set of jobs $L,M$, and $R$, respectively. The subschedules $S^*_I$ and $S^*_J$ are optimal schedules of $I$ and $J$, respectively.

    Consider the two schedules $S^{I \leftarrow J} = S_L \oplus S^*_J \oplus S^*_I \oplus S_M \oplus S_R$
    and $S^{I \rightarrow J} = S_L \oplus S_M \oplus S^*_J \oplus S^*_I \oplus S_R$ that are obtained from $S$ by moving $S^*_J$ directly before $S^*_I$ and moving $S^*_I$ directly after $S^*_J$, respectively.\footnote{Note that the order of $S^*_I$ and $S^*_J$ is changed, see \cref{fig:subscheduleorder:consistency}.}

    If $I \succeq J$ then $b \braces{S^{I \leftarrow J}} \le b(S)$ or $b \braces{S^{I \rightarrow J}} \le b(S)$.
    \begin{proof}
    %

      We use \Cref{def:budgetreturnorder:budget} and compare the budgets of the schedules $S^{I \leftarrow J}$ and $S^{I \rightarrow J}$ to $S$ by looking at the costs of corresponding prefixes.
      For $j \in N$, we denote by  $P_j$, $P^{I \leftarrow J}_j$, and $P^{I \rightarrow J}_j$ the prefix of $S$, $S^{I \leftarrow J}$, and $S^{I \rightarrow J}$ that ends with job $j$, respectively.
      As $S^*_I$ and $S^*_J$ are optimal schedules for $I$ and $J$, we have $b \braces{S^*_I} = b(I)$ and $b \braces{S^*_J} = b(J)$.
      Note that for $j \in L \cupdot R$ the costs of the prefixes do not change, i.e.
      $ c\braces{P^{I \leftarrow J}_j} = c\braces{P^{I \rightarrow J}_j} = c(P_j) \le b(S) $ by definition of the budget.

      We compare other prefixes depending on the cost of $I$ and $J$ using \Cref{obs:budgetreturnconcat}.
        \begin{enumerate}[align=left,leftmargin=*,font=\itshape]
            \item [Case $c(I), c(J) < 0$:]

            From $I \succeq J$, it follows that $b(I) \ge b(J)$ and thus $b \braces{S^*_I} \ge b \braces{S^*_J}$.
            \begin{align*}
                \text{For } j \in J &:& c\braces{P^{I \leftarrow J}_j} &\le c(S_L) + b(S^*_J) \le c(S_L) + b(S^*_I) \le b(S) & \\
                \text{For } j \in  I \cupdot M &:& c\braces{P^{I \leftarrow J}_j} &= c(P_j) + c(S^*_J) < c(P_j) \le b(S) &
            \end{align*}
            So $b \braces{S^{I \leftarrow J}} \le b(S)$.
            \item [Case $c(I) \ge 0 > c(J)$ and $c(M) \ge 0$:]

            \begin{align*}
                \text{For } j \in J &:& c \braces{P^{I \leftarrow J}_j} &= c(P_j) - c(S^*_I \oplus S_M) \le c(P_j) \le b(S) & \\
                \text{For } j \in I \cupdot M &:& c \braces{P^{I \leftarrow J}_j} &= c(P_j) + c(S^*_J) < c(P_j) \le b(S) &
            \end{align*}
           So $b \braces{S^{I \leftarrow J}} \le b(S)$.
            \item [Case $c(I) \ge 0 > c(J)$ and $c(M) < 0$:]
            \begin{align*}
                \text{For } j \in I &:& c \braces{P^{I \rightarrow J}_j} &= c(P_j) + c(S_M \oplus S^*_J) < c(P_j) \le b(S) &\\
                \text{For } j \in M \cupdot J &:& c \braces{P^{I \rightarrow J}_j} &= c(P_j) - c(S^*_I) \le c(P_j) \le b(S) &
            \end{align*}
            So $b \braces{S^{I \rightarrow J}} \le b(S)$.
            \item [Case $c(I), c(J) \ge 0$:]

            From $I \succeq J$, it follows that $r(I) \ge r(J)$ and thus $r(S^*_I) \ge r(S^*_J)$.
            \begin{align*}
                \text{For } j \in I &:& c \braces{P^{I \rightarrow J}_j} &\le c(N) - c(S_R) - r(S^*_I) \le c(N) - c(S_R) - r(S^*_J) = \max_{i \in J}  c(P_i) \leq b(S) &\\
                \text{For } j \in M \cupdot J &:& c \braces{P^{ I \rightarrow J}_j} &= c(P_j) - c(S^*_I) \le c(P_j) \le b(S) &
            \end{align*}
            So $b \braces{S^{I \rightarrow J}} \le b(S)$.
        \end{enumerate}

        In any case, $S^{I \leftarrow J}$ or $S^{I \rightarrow J}$ is not worse than $S$.
    \end{proof}
\end{lemma}

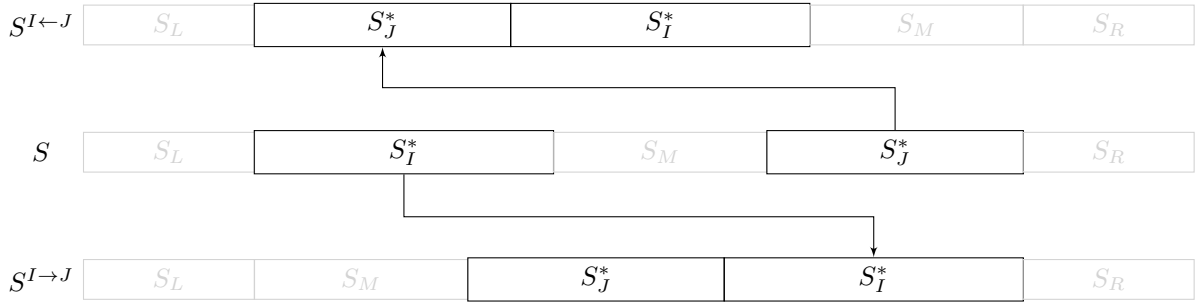
\begin{figure}
    \centering
    \begin{tikzpicture}[x=32pt, y=16pt, start chain=1, start chain=2, start chain=3, node distance=-.5pt]

        \tikzset{
            subschedule/.style={draw},
            gray/.style={draw=lightgray,text=lightgray}
        }

        \node [on chain=2,minimum width=32pt] at (0, -3) {$S^{I\rightarrow J}$};
        \node [on chain=2,subschedule,gray,minimum width= 64pt] {$S_L$};
        \node [on chain=2,subschedule,gray,minimum width= 80pt] {$S_M$};
        \node [on chain=2,subschedule,     minimum width= 96pt] {$S^*_J$  };
        \node [on chain=2,subschedule,     minimum width=112pt] (Srij) {$S^*_I$  };
        \node [on chain=2,subschedule,gray,minimum width= 64pt] {$S_R$  };

        \node [on chain=1,minimum width=32pt] at (0, 0) {$S$};
        \node [on chain=1,subschedule,gray,minimum width= 64pt] {$S_L$};
        \node [on chain=1,subschedule,     minimum width=112pt] (Sij) {$S^*_I$  };
        \node [on chain=1,subschedule,gray,minimum width= 80pt] {$S_M$};
        \node [on chain=1,subschedule,     minimum width= 96pt] (Skl) {$S^*_J$  };
        \node [on chain=1,subschedule,gray,minimum width= 64pt] {$S_R$};

        \node [on chain=3,minimum width=32pt] at (0, 3) {$S^{I \leftarrow J}$};
        \node [on chain=3,subschedule,gray,minimum width= 64pt] {$S_L$};
        \node [on chain=3,subschedule,     minimum width= 96pt] (Slkl) {$S^*_J$  };
        \node [on chain=3,subschedule,     minimum width=112pt] {$S^*_I$  };
        \node [on chain=3,subschedule,gray,minimum width= 80pt] {$S_M$};
        \node [on chain=3,subschedule,gray,minimum width= 64pt] {$S_R$};

        \draw[>=latex', ->] (Sij.south) -- ++(0,-1) -| (Srij.north);
        \draw[>=latex', ->] (Skl.north) -- ++(0, 1) -| (Slkl.south);

    \end{tikzpicture}
    \caption{Construction of $S^{I \leftarrow J}$ (top) and $S^{I \rightarrow J}$ (bottom) from \Cref{lem:consistency}.}
    \label{fig:subscheduleorder:consistency}
\end{figure}

The above lemma states that given a partition of the jobs into blocks, all of which are scheduled optimally and contiguously, the order of the block schedules following the cbr-preorder incurs the lowest budget.
We now present a generic algorithm that iteratively chooses a minimal ideal w.r.t.~the cbr-preorder as the next subset of jobs in the schedule. This subset is scheduled optimally, and is removed from consideration.
The algorithm is stated in \Cref{alg:minbudget:generic}.

\begin{algorithm}
	\setstretch{1.25}
	\medskip
	\KwInput{instance $(N, \unlhd, c)$ of \minbudget}
	\KwOutput{optimal schedule $S$ for $(N, \unlhd, c)$}
	\medskip
    $S \leftarrow$ empty schedule \;
    \While{$\card{S} < \card{N}$}{
        Pick an inclusion-maximal minimal ideal $I$ of $\unlhd_{N \setminus S}$ with respect to $\preceq$ \;
        Choose an optimal schedule $S^*_I$ of $\unlhd_I$ \;
        $S \leftarrow S \oplus S^*_I$ \;
    }
    \Return $S$ \;
	\medskip
	\caption{Generic Algorithm for Budget Minimization}
	\label{alg:minbudget:generic}
\end{algorithm}

We will show that the schedule computed by \Cref{alg:minbudget:generic} is indeed in line with the cbr-preorder in \Cref{sec:optimalityofalgorithm}.
Note that it is not clear how to pick the ideal $I$ or how to compute an optimal schedule~$S^*_I$ in each iteration.
However, we use \cref{alg:minbudget:generic} mainly to prove that there always exists an optimal schedule of a certain structure.
Therefore, we do not care if and how we can efficiently compute $I$ and its optimal schedule $S^*_I$.
By choosing only ideals of the remaining jobs, we immediately see that the algorithm returns a feasible schedule.

\begin{observation}\label{obs:algorithm:feasible}
The schedule returned by \Cref{alg:minbudget:generic} is feasible.
\end{observation}

The remainder of this section is dedicated to showing optimality of the computed schedule.

\subsection{Irreducible Intervals}\label{sec:irreducibleintervals}

In the previous subsection, we showed that schedules following the cbr-preorder of the instance have small budget.
But we did not consider the feasibility of the occurring schedules.
In particular, the schedules $S^{I \rightarrow J}$ and $S^{I \leftarrow J}$ constructed in \Cref{lem:consistency} may not be feasible.
In this section, we will derive the necessary tools to show that for the right choice of the partition, there is a feasible schedule that is in line with the cbr-preorder.
This will ultimately show that \Cref{alg:minbudget:generic} computes an optimal solution.

\begin{definition}[Irreducible Intervals]
\label{def:irreducibility}
    Let $(N,\unlhd,c)$ be an instance of \minbudget.
        An interval $I \subseteq N$ of $\unlhd$ is called \emph{irreducible} if $I$ is a minimal ideal of itself w.r.t.~$\preceq$, i.e.,~$J \succeq I$ for all ideals $J \subseteq I$ of $\unlhd_I$.
\end{definition}

The main intuition of the notion of irreducibility is that we would always either schedule an irreducible interval completeley, or not at all.
That is, it does not make sense w.r.t.~the total budget to preempt an irreducible interval.
Note that the definition of $I$ being irreducible depends only on the interval $I$ itself, and not on the set $N \setminus I$.
That is if $I \subseteq N$ is an irreducible interval w.r.t.~$\unlhd$, and $N'$ is a set satisfying $I \subseteq N' \subseteq N$, then $I$ is still irreducible w.r.t.~$\unlhd_{N'}$.
The minimal ideals of $\unlhd$ w.r.t.~the cbr-preorder are irreducible, i.e.,~\cref{alg:minbudget:generic} picks irreducible intervals in each iteration.
We get the following relations of cost, budget, and return of an irreducible interval and its ideals and filters.

\begin{lemma}[Properties of Irreducible Intervals]
\label{lem:irreducibleinterval}
    Let $(N, \unlhd, c)$ be an instance of \minbudget~and $I \subseteq N$ be an irreducible interval of $\unlhd$.
    Then
    \begin{lemmaenum}
    %
    %
        \item \label{lem:irreducibleinterval:mincostideal}
        $c(J) \ge \min \set{0, c(I)}$ for all ideals $J \subseteq I$ of $\unlhd_I$,
        \item \label{lem:irreducibleinterval:maxcostfilter}
        $c(F) \le \max \set{0, c(I)}$ for all filters $F \subseteq I$ of $\unlhd_I$,
        \item \label{lem:irreducibleinterval:ideal}
        $c(J) \ge 0$ for all ideals $J \subseteq I$ of $\unlhd_I$ with $b(J) < b(I)$,
        \item \label{lem:irreducibleinterval:filter}
        $c(F) \le 0$ for all filters $F \subseteq I$ of $\unlhd_I$ with $r(F) > r(I)$, and
        \item \label{lem:irreducibleinterval:minreturnideal}
        $r(J) \ge r(I)$ for all ideals $J \subseteq I$ of $\unlhd_I$.
    \end{lemmaenum}
    \begin{proof}
       Let $S^*_I$ be an optimal schedule of $(I,\unlhd_I,c)$, i.e.,~$b \braces{S^*_I} = b(I)$.
        \begin{lemmaenum}
        %
        %
            \item
            Let $J \subseteq I$ be an inclusion-minimal ideal of $\unlhd_I$ that minimizes $c(J)$.
            Then, $c(J^\prime) \ge 0$ for every ideal $J^\prime \subseteq I \setminus J$ of $\unlhd_{I \setminus J}$.
            Assume for a contradiction that $c(J) < \min \set{0, c(I)}$.
            Then irreducibility of $I$ implies $c(I) < 0$ and $b(J) \geq b(I)$.
            Consider the schedule $S := S^*_I \cap J$ to be the subschedule of $S^*_I$ on the jobs in $J$.
            $S$ is a feasible schedule for $J$ as $J$ is an ideal.
            For a prefix $P^*$ of $S^*_I$ and the corresponding prefix $P := S \cap P^*$ of $S$, it holds $c(P) = c \braces{P^*} - c \braces{P^* \setminus J} \le c \braces{P^*}$ since the jobs in $P^* \setminus J$ form an ideal of $\unlhd_{I \setminus J}$.
            Hence, $b(J) \le b(S) \le b \braces{S^*_I} = b(I)$ holds.
            So $b(J) = b(I)$, and $r(J) = c(J) - b(J) < c(I) - b(I) = r(I)$.
            This yields $J \prec I$ which contradicts the irreducibility of $I$.
            \item
            Let $F \subseteq I$ be a filter of $\unlhd_I$. Then, $I \setminus F$ is an ideal of $\unlhd_I$.
            From \Cref{lem:irreducibleinterval:mincostideal}, we have $c(F) = c(I) - c(I \setminus F) \le c(I) - \min \set{0, c(I)} = \max \set{0, c(I)}$.
            \item
            Let $J \subseteq I$ be an ideal of $\unlhd_I$ with $b(J) < b(I)$.
            Then $c(J) < 0$ would imply $J \prec I$, which contradicts the irreduciblity of $I$.
            \item
            Let $F \subseteq I$ be a filter of maximum cost of $\unlhd_I$ with $r(F) > r(I)$.
            Suppose that $c(F) > 0$.
            By \Cref{lem:irreducibleinterval:maxcostfilter}, it holds $c(I) > 0$.
            Note that $I \setminus F$ is an ideal of $\unlhd_I$ with $c(I \setminus F) = c(I) - c(F) < c(I)$.
            If $c(I \setminus F) < 0$, we get a contradiction to the irreducibility of $I$, so $0 \le c(I \setminus F) < c(I)$.

            Let $S^*_{I \setminus F}$ and $S^*_F$ be optimal schedules of $I \setminus F$ and $F$, respectively.
            Consider the schedule $S = S^*_{I \setminus F} \oplus S^*_F$ which is feasible for $I$.
            In particular, $b(S) \geq b(I)$.
            For a prefix $P$ of $S$ with $P \cap F \ne \emptyset$, it holds $c(P) \leq c(I) - r(F) < c(I) - r(I) = b(I) \leq b(S)$.
            So $b(S) > b(S^*_F) + c(S^*_{I \setminus F})$, and $b(S) = \max\{b(S^*_{I \setminus F}), b(S^*_F) + c(S^*_{I \setminus F})\} = b(I \setminus F)$ by \Cref{obs:budgetreturncost:composition:budget}. Together with $b(S) \geq b(I)$ we obtain
            \begin{equation*}
            r(I \setminus F) = c(I \setminus F) - b(I \setminus F) = c(I \setminus F) - b(S) < c(I) - b(I) = r(I).
            \end{equation*}
            However, then $I \setminus F \prec I$, a contradiction to the irreducibility of $I$. The claim follows by our choice of $F$.
            \item
            Let $J \subseteq I$ be an ideal of $\unlhd_I$.
            If $c(I) \ge 0$, then $c(J) \ge 0$ and $r(J) \ge r(I)$ follows from the irreducibility of $I$.
            Hence, we can assume that $c(I) < 0$.

            Let $S := S^*_I \cap J$ be the subschedule of $S^*_I$ on the jobs in $J$. Note that $S$ is feasible for $J$.
            Let $Q^*$ be a suffix of $S^*_I$ and $Q = S \cap Q$ the corresponding suffix of $S$.
            Then $c(Q) = c \braces{Q^*} - c \braces{Q^* \setminus J} \ge c \braces{Q^*}$ follows from \Cref{lem:irreducibleinterval:maxcostfilter}, since $I \setminus J$, and thus $Q^* \setminus J$ is a filter of $\unlhd_I$.
            Hence, $r(J) \ge r(S) \ge r(S^*_I) = r(I)$ as claimed. \qedhere
        \end{lemmaenum}
    \end{proof}
\end{lemma}

The above lemma formalizes the intuition of irreducible intervals.
If we scheduled a part of the irreducible interval, but did not yet reach the budget, we know by \Cref{lem:irreducibleinterval:ideal} that it is better to not schedule this part at all.
If we reached the budget of the interval, \Cref{lem:irreducibleinterval:filter} yields that scheduling the rest of the interval will only help to decrease the overall budget.
It turns out that, given a schedule, we can contiguously schedule the jobs of an irreducible interval, without increasing the budget of a schedule.
Note that the resulting schedule might not be feasible.
We again use the notation $i \mathbin{S} j$, if $i$ precedes $j$ in the schedule $S$ (see \Cref{sec:hardness}).

\begin{lemma}[Non-Preemption and Optimal Subschedules of Irreducible Intervals]
\label{lem:nonpreemption}
    Let $(N, \unlhd, c)$ be an instance of \minbudget~and let $S$ be a feasible schedule.
    Let $I \subseteq N$ be an irreducible interval of $\unlhd$ and $S^*_I$ be an optimal schedule of $(I,\unlhd_I,c)$, i.e.,~$b \braces{S^*_I} = b(I)$.
    Set $L_I := \set{j \in N \colon j \mathbin{S} i \text{ for all } i \in I}$ and $R_I := \set{j \in N \colon i \mathbin{S} j \text{ for all } i \in I}$.

    Then there exists a prefix $P$ and a suffix $Q$ of $S = P \oplus Q$ with $L_I \subseteq P$ and $R_I \subseteq Q$ such that the schedule
    $S^\prime := \braces{P \setminus I} \oplus S^*_I \oplus \braces{Q \setminus I }$ satisfies $b(S^\prime) \le b(S)$.
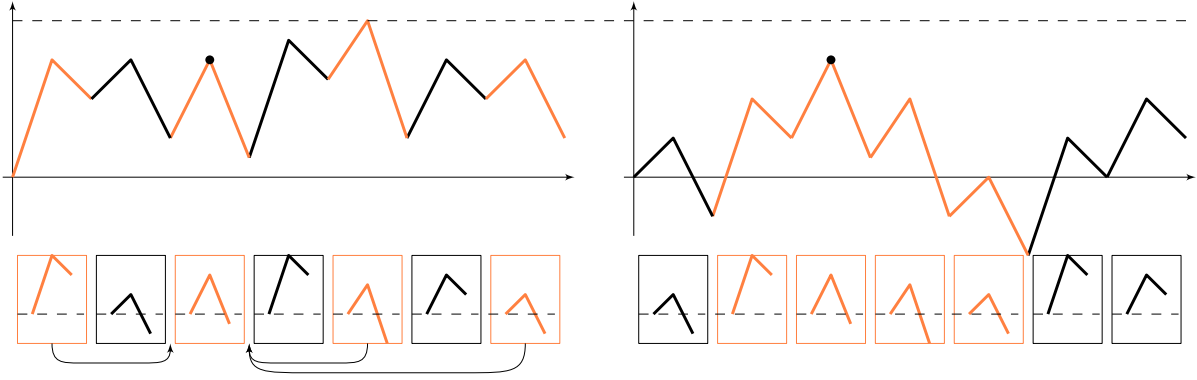
\begin{figure}
	\resizebox{\textwidth}{!}{
	\begin{tikzpicture}[ipe stylesheet]
  \draw[shift={(64, 728.001)}, yscale=1.3333, ->]
    (0, 0)
     -- (0, 72);
  \draw[ipe pen fat]
    (160, 760)
     -- (176, 808)
     -- (192, 792);
  \draw[ipe pen fat]
    (224, 768)
     -- (240, 800)
     -- (256, 784);
  \draw[rgb color={draw=1 0.502 0.251}, ipe pen fat]
    (128, 768)
     -- (144, 800)
     -- (160, 760);
  \draw[rgb color={draw=1 0.502 0.251}, ipe pen fat]
    (192, 792)
     -- (208, 816)
     -- (224, 768);
  \draw[rgb color={draw=1 0.502 0.251}, ipe pen fat]
    (256, 784)
     -- (272, 800)
     -- (288, 768);
  \draw[shift={(98, 720)}, xscale=0.875, yscale=2.25]
    (0, 0) rectangle (32, -16);
  \draw[shift={(162, 720)}, xscale=0.875, yscale=2.25]
    (0, 0) rectangle (32, -16);
  \draw[shift={(226, 720)}, xscale=0.875, yscale=2.25]
    (0, 0) rectangle (32, -16);
  \draw[shift={(130, 720)}, xscale=0.875, yscale=2.25, rgb color={draw=1 0.502 0.251}]
    (0, 0) rectangle (32, -16);
  \draw[shift={(194, 720)}, xscale=0.875, yscale=2.25, rgb color={draw=1 0.502 0.251}]
    (0, 0) rectangle (32, -16);
  \draw[shift={(258, 720)}, xscale=0.875, yscale=2.25, rgb color={draw=1 0.502 0.251}]
    (0, 0) rectangle (32, -16);
  \draw[-{>[ipe arrow tiny]}]
    (208, 684)
     .. controls (208, 676) and (204, 676) .. (196.6667, 676)
     .. controls (189.3333, 676) and (178.6667, 676) .. (171.3333, 676)
     .. controls (164, 676) and (160, 676) .. (160, 684);
  \draw[-{>[ipe arrow tiny]}]
    (272, 684)
     .. controls (272, 672) and (266, 672) .. (248.3333, 672)
     .. controls (230.6667, 672) and (201.3333, 672) .. (183.6667, 672)
     .. controls (166, 672) and (160, 672) .. (160, 684);
  \draw[shift={(168, 696)}, scale=0.5, ipe pen fat]
    (0, 0)
     -- (16, 48)
     -- (32, 32);
  \draw[shift={(232, 696)}, scale=0.5, ipe pen fat]
    (0, 0)
     -- (16, 32)
     -- (32, 16);
  \draw[shift={(136, 696)}, scale=0.5, rgb color={draw=1 0.502 0.251}, ipe pen fat]
    (0, 0)
     -- (16, 32)
     -- (32, -8);
  \draw[shift={(200, 696)}, scale=0.5, rgb color={draw=1 0.502 0.251}, ipe pen fat]
    (0, 0)
     -- (16, 24)
     -- (32, -24);
  \draw[shift={(264, 696)}, scale=0.5, rgb color={draw=1 0.502 0.251}, ipe pen fat]
    (0, 0)
     -- (16, 16)
     -- (32, -16);
  \draw[ipe dash dashed]
    (98, 696)
     -- (125, 696);
  \draw[ipe dash dashed]
    (130, 696)
     -- (157, 696);
  \draw[ipe dash dashed]
    (162, 696)
     -- (189, 696);
  \draw[ipe dash dashed]
    (194, 696)
     -- (221, 696);
  \draw[ipe dash dashed]
    (226, 696)
     -- (253, 696);
  \draw[ipe dash dashed]
    (257, 696)
     -- (284, 696);
  \draw[shift={(316, 728.001)}, yscale=1.3333, ->]
    (0, 0)
     -- (0, 72);
  \draw[shift={(312, 752)}, xscale=1.7059, ->]
    (0, 0)
     -- (136, 0);
  \draw[ipe pen fat]
    (476, 720)
     -- (492, 768)
     -- (508, 752);
  \draw[ipe pen fat]
    (508, 752)
     -- (524, 784)
     -- (540, 768);
  \draw[rgb color={draw=1 0.502 0.251}, ipe pen fat]
    (380, 768)
     -- (396, 800)
     -- (412, 760);
  \draw[rgb color={draw=1 0.502 0.251}, ipe pen fat]
    (412, 760)
     -- (428, 784)
     -- (444, 736);
  \draw[rgb color={draw=1 0.502 0.251}, ipe pen fat]
    (444, 736)
     -- (460, 752)
     -- (476, 720);
  \draw[shift={(318, 720)}, xscale=0.875, yscale=2.25]
    (0, 0) rectangle (32, -16);
  \draw[shift={(478, 720)}, xscale=0.875, yscale=2.25]
    (0, 0) rectangle (32, -16);
  \draw[shift={(510, 720)}, xscale=0.875, yscale=2.25]
    (0, 0) rectangle (32, -16);
  \draw[shift={(382, 720)}, xscale=0.875, yscale=2.25, rgb color={draw=1 0.502 0.251}]
    (0, 0) rectangle (32, -16);
  \draw[shift={(414, 720)}, xscale=0.875, yscale=2.25, rgb color={draw=1 0.502 0.251}]
    (0, 0) rectangle (32, -16);
  \draw[shift={(446, 720)}, xscale=0.875, yscale=2.25, rgb color={draw=1 0.502 0.251}]
    (0, 0) rectangle (32, -16);
  \draw[shift={(484, 696)}, scale=0.5, ipe pen fat]
    (0, 0)
     -- (16, 48)
     -- (32, 32);
  \draw[shift={(516, 696)}, scale=0.5, ipe pen fat]
    (0, 0)
     -- (16, 32)
     -- (32, 16);
  \draw[shift={(388, 696)}, scale=0.5, rgb color={draw=1 0.502 0.251}, ipe pen fat]
    (0, 0)
     -- (16, 32)
     -- (32, -8);
  \draw[shift={(420, 696)}, scale=0.5, rgb color={draw=1 0.502 0.251}, ipe pen fat]
    (0, 0)
     -- (16, 24)
     -- (32, -24);
  \draw[shift={(452, 696)}, scale=0.5, rgb color={draw=1 0.502 0.251}, ipe pen fat]
    (0, 0)
     -- (16, 16)
     -- (32, -16);
  \draw[ipe dash dashed]
    (318, 696)
     -- (345, 696);
  \draw[ipe dash dashed]
    (382, 696)
     -- (409, 696);
  \draw[ipe dash dashed]
    (478, 696)
     -- (505, 696);
  \draw[ipe dash dashed]
    (414, 696)
     -- (441, 696);
  \draw[ipe dash dashed]
    (510, 696)
     -- (537, 696);
  \draw[ipe dash dashed]
    (445, 696)
     -- (472, 696);
  \pic
     at (396, 800) {ipe disk};
  \pic
     at (144, 800) {ipe disk};
  \draw[shift={(350, 720)}, xscale=0.875, yscale=2.25, rgb color={draw=1 0.502 0.251}]
    (0, 0) rectangle (32, -16);
  \draw[ipe dash dashed]
    (349, 696)
     -- (376, 696);
  \draw[shift={(60, 752)}, xscale=1.7059, ->]
    (0, 0)
     -- (136, 0);
  \draw[shift={(66, 720)}, xscale=0.875, yscale=2.25, rgb color={draw=1 0.502 0.251}]
    (0, 0) rectangle (32, -16);
  \draw[ipe dash dashed]
    (66, 696)
     -- (93, 696);
  \draw[rgb color={draw=1 0.502 0.251}, ipe pen fat]
    (64, 752)
     -- (80, 800)
     -- (96, 784);
  \draw[ipe pen fat]
    (96, 784)
     -- (112, 800)
     -- (128, 768);
  \draw[shift={(72, 696)}, scale=0.5, rgb color={draw=1 0.502 0.251}, ipe pen fat]
    (0, 0)
     -- (16, 48)
     -- (32, 32);
  \draw[shift={(104, 696)}, scale=0.5, ipe pen fat]
    (0, 0)
     -- (16, 16)
     -- (32, -16);
  \draw[shift={(324, 696)}, scale=0.5, ipe pen fat]
    (0, 0)
     -- (16, 16)
     -- (32, -16);
  \draw[shift={(356, 696)}, scale=0.5, rgb color={draw=1 0.502 0.251}, ipe pen fat]
    (0, 0)
     -- (16, 48)
     -- (32, 32);
  \draw[rgb color={draw=1 0.502 0.251}, ipe pen fat]
    (348, 736)
     -- (364, 784)
     -- (380, 768);
  \draw[ipe pen fat]
    (316, 752)
     -- (332, 768)
     -- (348, 736);
  \draw[{<[ipe arrow tiny]}-]
    (128, 684)
     .. controls (128, 676) and (124, 676) .. (116.6667, 676)
     .. controls (109.3333, 676) and (98.6667, 676) .. (91.3333, 676)
     .. controls (84, 676) and (80, 676) .. (80, 684);
  \draw[ipe dash dashed]
    (64, 816)
     -- (544, 816);
\end{tikzpicture}}
	\caption{Illustration to \Cref{lem:nonpreemption}. The left figure depicts a schedule $S$, where the orange part forms an irreducible interval $I$ of $\unlhd$. The $\cdot$ describes the point at which $I$ attains its budget. On the right, the schedule $S^\prime := \braces{P \setminus I} \oplus S^*_I \oplus \braces{Q \setminus I }$, where $I$ is scheduled contiguously, is depicted.}
	\label{Fig:nonpreemption}
\end{figure}
    \begin{proof}
        We will define the prefix $P$ and the suffix $Q$ with $S = P \oplus Q$ such that $c(P^\prime \cap I) \ge 0$ for all prefixes $P^\prime \subseteq P$ and $c(Q^\prime \cap I) \le 0$ for all suffixes $Q^\prime \subseteq Q$.
        That is the value $b(I)$ within the jobs $S \cap I$ is attained at the (end of) last job in $P$, see \Cref{Fig:nonpreemption}.
        The definition of $P$ and $Q$ depends on the sign of $c(I)$.

        \emph{Case $c(I) < 0$:}
        Let $P \subseteq S$ be the inclusion-minimal prefix of $S$ such that $b(P \cap I) \ge b(I)$ and $L_I \subseteq P$.
        Set $Q := S \setminus P$ to be the corresponding suffix such that $S = P \oplus Q$.
        Such a prefix and suffix exist since $S \cap I$ is a feasible schedule for $\unlhd_I$, in particular $b(S \cap I) \ge b(I)$.
        Then $c(P \cap I) = b(P \cap I) \ge b(I) \ge 0$ by the inclusion-minimality of $P$.
        For all proper prefixes $P^\prime$ of $P$, it holds $b(P^\prime \cap I) < b(I)$ and $c(P^\prime \cap I) \geq 0$ (\Cref{lem:irreducibleinterval:ideal}).
        For all suffixes $Q^\prime$ of $Q$, it holds $c(Q^\prime \cap I) \leq 0$ (\Cref{lem:irreducibleinterval:maxcostfilter}).

        \emph{Case $c(I) \ge 0$:}
        Let $Q \subseteq S$ be the inclusion-minimal suffix of $S$ such that $r(Q \cap I) \le r(I)$ and $R_I \subseteq Q$.
        Set $P := S \setminus Q$ to be the corresponding prefix such that $S = P \oplus Q$.
        Such a prefix and suffix exist since $S \cap I$ is a feasible schedule for $\unlhd_I$, in particular $r(S \cap I) \le r(I)$.
        As $Q$ is chosen to be inclusion-minimal, $c(Q \cap I) = r(Q \cap I) \le r(I) \le 0$.
        Further, $c(P \cap I) = c(I) - c(Q \cap I) = c(I) - r(Q \cap I) \ge c(I) - r(I) = b(I)$.
        For all prefixes $P^\prime$ of $P$, it holds $c(P^\prime \cap I) \geq 0$ (\Cref{lem:irreducibleinterval:mincostideal}).
        For all proper suffixes $Q^\prime$ of $Q$, it holds $r(Q^\prime \cap I) > r(I)$ and $c(Q^\prime \cap I) \leq 0$ (\Cref{lem:irreducibleinterval:filter}).

        In either case, we obtain the following costs for prefixes of $S^\prime := (P \setminus I) \oplus S^*_I \oplus (Q \setminus I)$.
        For $j \in P \setminus I$, let $P_j$ and $P^\prime_j$ be the prefixes of $S$ and $S^\prime$ that contain all jobs until $j$, respectively. Note that $P_j$ is a prefix of $P$.
        Hence $c(P^\prime_j) = c(P_j) - c(P_j \cap I) \leq c(P_j)$.
        Similarly, for $j \in Q \setminus I$, the subschedule $S \setminus P_j$ is a suffix of $Q$ and, hence,
        $c(P^\prime_j) = c(P_j) + c\braces{(S \setminus P_j) \cap I} \leq c(P_j)$.
        Finally, for $j \in I$, and with $P^*_j$ being the prefix of $S^*_I$ that contains all jobs until $j$, it holds
    	\begin{equation*}
            c(P^\prime_j) = c(P \setminus I) + c \braces{P^*_j} \leq c(P \setminus I) + b(I) \leq c(P \setminus I) + c(P \cap I) = c(P) .
    	\end{equation*}
    	So in total, $b(S^\prime) \leq b(S)$.
    \end{proof}
\end{lemma}

\subsection{Schedules in Increasing Irreducible Structure}\label{sec:optimalityofalgorithm}

Recall that the minimal ideals that \Cref{alg:minbudget:generic} chooses are irreducible intervals.
In fact, the algorithm computes a schedule of a certain structure that is consistent with the cbr-preorder.
The structure we are aiming for is defined as follows.

\begin{definition}[Schedules in Increasing Irreducible Structure]
\label{def:universality}
    Let $(N,\unlhd,c)$ be an instance of \minbudget.
    A feasible schedule $S = S_1 \oplus \ldots \oplus S_k$ for $(N, \unlhd, c)$ is said to be in \emph{increasing irreducible structure} if there are disjoint sets $I_1, \ldots, I_k \subseteq N$ such that
    \begin{lemmaenum}
      \item $I_1, \ldots, I_k$ are irreducible intervals of $\unlhd$,
      \item $S_i$ is an optimal schedule of $I_i$ for all $i \in \set{1, \ldots, k}$, and
      \item $I_i \preceq I_j$ for all $1 \le i < j \le k$.
    \end{lemmaenum}
\end{definition}

For a schedule $S$ in increasing irreducible structure given as $S = S_1 \oplus \ldots \oplus S_k$ we call irreducible intervals $I_1, \ldots, I_k$ of $\unlhd$ \emph{corresponding irreducible intervals} for $S$ if $S$ together with $I_1, \ldots , I_k$ fulfills \Cref{def:universality}.

Recall from the comment to \Cref{def:irreducibility} that irreducibility is a local notion, in the sense that it does not depend on other parts of the partially ordered set than the interval it refers to.
For a schedule $S$ in increasing irreducible structure, there may be different ways of partitioning the jobs into irreducible intervals, see \cref{fig:irreduciblestructure:notunique} for an example.

The next lemma shows that any schedule in increasing irreducible structure is indeed optimal.

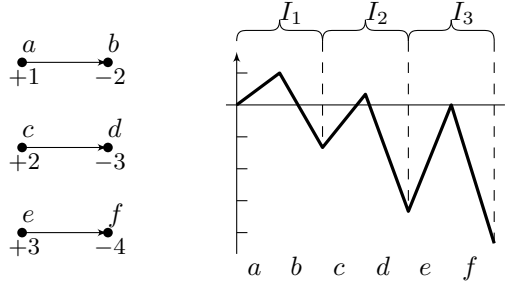
\begin{figure}
	\centering
	\begin{tikzpicture}[ipe stylesheet]
  \draw
    (176, 756)
     -- (180, 756);
  \draw
    (176, 732)
     -- (180, 732);
  \draw
    (176, 720)
     -- (180, 720);
  \draw[-{>[ipe arrow small]}]
    (176, 688)
     -- (176, 764);
  \pic
     at (96, 760) {ipe disk};
  \pic
     at (128, 760) {ipe disk};
  \draw[-{>[ipe arrow small]}]
    (96, 760)
     -- (128, 760);
  \node[ipe node]
     at (128, 764) {$b$};
  \node[ipe node]
     at (96, 764) {$a$};
  \node[ipe node, font=\small]
     at (122.902, 751.988) {$-2$};
  \node[ipe node, font=\small]
     at (90.902, 751.988) {$+1$};
  \pic
     at (96, 728) {ipe disk};
  \pic
     at (128, 728) {ipe disk};
  \draw[-{>[ipe arrow small]}]
    (96, 728)
     -- (128, 728);
  \node[ipe node]
     at (128, 732) {$d$};
  \node[ipe node]
     at (96, 732) {$c$};
  \node[ipe node, font=\small]
     at (122.902, 719.988) {$-3$};
  \node[ipe node, font=\small]
     at (90.902, 719.988) {$+2$};
  \pic
     at (96, 696) {ipe disk};
  \pic
     at (128, 696) {ipe disk};
  \draw[-{>[ipe arrow small]}]
    (96, 696)
     -- (128, 696);
  \node[ipe node]
     at (128, 700) {$f$};
  \node[ipe node]
     at (96, 700) {$e$};
  \node[ipe node, font=\small]
     at (122.902, 687.988) {$-4$};
  \node[ipe node, font=\small]
     at (90.902, 687.988) {$+3$};
  \draw
    (176, 708)
     -- (180, 708);
  \draw
    (176, 696)
     -- (180, 696);
  \node[ipe node]
     at (180, 680) {$a$};
  \node[ipe node]
     at (196, 680) {$b$};
  \node[ipe node]
     at (212, 680) {$c$};
  \node[ipe node]
     at (228, 680) {$d$};
  \node[ipe node]
     at (244, 680) {$e$};
  \node[ipe node]
     at (260, 680) {$f$};
  \draw
    (276, 744)
     -- (172, 744);
  \draw
    (176, 768)
     .. controls (176, 772) and (178, 772) .. (180.3333, 772)
     .. controls (182.6667, 772) and (185.3333, 772) .. (187.6667, 772)
     .. controls (190, 772) and (192, 772) .. (192, 776);
  \draw
    (192, 776)
     .. controls (192, 772) and (194, 772) .. (196.3333, 772)
     .. controls (198.6667, 772) and (201.3333, 772) .. (203.6667, 772)
     .. controls (206, 772) and (208, 772) .. (208, 768);
  \draw
    (208, 768)
     .. controls (208, 772) and (210, 772) .. (212.3333, 772)
     .. controls (214.6667, 772) and (217.3333, 772) .. (219.6667, 772)
     .. controls (222, 772) and (224, 772) .. (224, 776);
  \draw
    (224, 776)
     .. controls (224, 772) and (226, 772) .. (228.3333, 772)
     .. controls (230.6667, 772) and (233.3333, 772) .. (235.6667, 772)
     .. controls (238, 772) and (240, 772) .. (240, 768);
  \draw
    (240, 768)
     .. controls (240, 772) and (242, 772) .. (244.3333, 772)
     .. controls (246.6667, 772) and (249.3333, 772) .. (251.6667, 772)
     .. controls (254, 772) and (256, 772) .. (256, 776);
  \draw
    (256, 776)
     .. controls (256, 772) and (258, 772) .. (260.3333, 772)
     .. controls (262.6667, 772) and (265.3333, 772) .. (267.6667, 772)
     .. controls (270, 772) and (272, 772) .. (272, 768);
  \node[ipe node]
     at (192, 776) {$I_1$};
  \node[ipe node]
     at (224, 776) {$I_2$};
  \node[ipe node]
     at (256, 776) {$I_3$};
  \draw[ipe pen fat]
    (176, 744)
     -- (192, 756)
     -- (208, 728)
     -- (224, 748)
     -- (240, 704)
     -- (256, 744)
     -- (272, 692);
  \draw[ipe dash dashed]
    (208, 768)
     -- (208, 728);
  \draw[ipe dash dashed]
    (240, 768)
     -- (240, 704);
  \draw[ipe dash dashed]
    (272, 768)
     -- (272, 692);
\end{tikzpicture}
	\caption{The schedule on the right is in increasing irreducible structure, but there are different ways of partitioning the jobs into irreducible intervals. Let $I_1 := \set{a,b}, I_2 := \set{c,d}$, and $I_3 := \set{e,f}$. All these intervals are irreducible. $I_1 \protect\cupdot I_2 \protect\cupdot I_3$, $(I_1 \cup I_2) \protect\cupdot I_3$ and $(I_1 \cup I_2 \cup I_3)$ are feasible choices of partitioning the jobs into irreducible intervals such that the schedule is in increasing irreducible structure.
  }
	\label{fig:irreduciblestructure:notunique}
\end{figure}

\begin{lemma}[Increasing Irreducible Structure is Optimal]\label{lem:iisoptimal}
  Let $(N, \unlhd, c)$ be an instance of \minbudget.
  If $S$ is a schedule for $(N, \unlhd, c)$ in increasing irreducible structure, then $S$ is optimal, i.e.,~$b(S) = b(N)$.
  \begin{proof}
    Let  $S = S_1 \oplus \ldots \oplus S_k$ and $I_1,\dots,I_k$ be the corresponding irreducible intervals.
    Let $S^*$ be an optimal schedule of $(N, \unlhd, c)$.
    Applying \Cref{lem:nonpreemption} to $S^*$ repeatedly for $I_1,\dots,I_k$, we obtain a schedule $S'$ with $b(S') \leq b(S^*)$ where each interval $I_i$ is scheduled contiguously and in optimal order $S_i$ for all $i \in [k]$.
    (Note that the schedule $S'$ is not necessarily feasible.)
    Then applying \Cref{lem:consistency} repeatedly to the schedule $S'$, we obtain the schedule $S$ and observe that $b(S) \leq b(S') \leq b(S^*)$.
    Therefore, $S$ is an optimal schedule.
  \end{proof}
\end{lemma}


The next lemma explains the choice of picking inclusion-maximal minimal ideals w.r.t.~$\preceq$ in the iterations of the algorithm.
This choice together with the lemma implies that the intervals are non-decreasing w.r.t.~$\preceq$ as it is required in the increasing irreducible structure.

\begin{lemma}[Order of Consecutive Intervals]\label{lem:mergeconsecblocksfallsfalschrum}
  Let $(N, \unlhd, c)$ be an instance of \minbudget.
  Let $I \subseteq N$ be an irreducible ideal of $\unlhd$ and let $J \subseteq N \setminus I$ be an irreducible ideal of
  $\unlhd_{N \setminus I}$.

  If $I \succ J$, then $I \cupdot J \preceq I$ holds.
  \begin{proof}
    Let $S^*_I$, $S^*_J$, and $S^*$ be optimal schedules for $I$, $J$, and $I \cupdot J$, respectively.
    Since $I$ and $J$ are irreducible intervals of $\unlhd$, we can apply \Cref{lem:nonpreemption,lem:consistency} to $S^*$ for $I$ and $J$.
    From $I \succ J$, we obtain the (not necessarily feasible) schedule $S' := S^*_J \oplus S^*_I$ with $b(S') \le b(S^*)$.
    From \Cref{obs:budgetreturncost:composition:return}, it follows $r(S') \le r(S^*_I)$.
    For the return of $I \cupdot J$ we obtain $r(I \cupdot J) = c(S^*) - b(S^*) \le c(S') - b(S') = r(S') \le r(S^*_I) = r(I)$.
    We distinguish between two cases w.r.t.~the cost of $I$:

    \emph{Case $c(I) \ge 0$:} From $r(I \cupdot J) \le r(I)$, we immediately have $I \cupdot J \preceq I$.

    \emph{Case $c(I) < 0$:} From $I \succ J$, we also have $c(J) < 0$ and $b(J) \le b(I)$.
    So the budget of $I \cupdot J$ satisfies
    $b(I \cupdot J) \le b(S^*_I \oplus S^*_J) = \max \set{ b(S^*_I), c(S^*_I) + b(S^*_J) } \le \max \set{b(S^*_I), b(S^*_J)} = \max \set{b(I), b(J)} \le b(I)$.

    Summarizing, it holds $r(I \cupdot J) \le r(I)$ and $b(I \cupdot J) \le b(I)$.
    As $c(I) < 0$, we thus obtain $I \cupdot J \preceq I$.
  \end{proof}
\end{lemma}

We are now ready to prove that \Cref{alg:minbudget:generic} computes a schedule in increasing irreducible structure. In particular, there always exists a schedule in increasing irreducible structure.

\begin{theorem}[Correctness of \Cref{alg:minbudget:generic}]
\label{thm:minbudget:generic:correctness}
    Let $(N,\unlhd,c)$ be an instance of \minbudget.
    \Cref{alg:minbudget:generic} computes a schedule for $(N,\unlhd,c)$ in increasing irreducible structure.
    In particular, this schedule is optimal.
    \begin{proof}
    		Let  $S = S^*_1 \oplus \ldots \oplus S^*_k$ be the schedule returned by \Cref{alg:minbudget:generic}, and let $I_1,\dots,I_k$ be the corresponding ideals chosen by the algorithm.
        By definition, minimal ideals w.r.t.~$\preceq$ are irreducible.
        For $S$ to be in increasing irreducible structure, we need $I_i \preceq I_j$ for all $1 \leq i < j \leq k$.

        We show that for every $i \in \set{1, \ldots, k-1}$ it holds $I_i \preceq I_{i+1}$.
        Fix $i \in \set{1, \dots, k-1}$.
        From the choices of the algorithm, we have that $I_i$ is an irreducible ideal of $\unlhd_{N \setminus \braces{I_1 \cup \dots \cup I_{i-1}}}$ and $I_{i+1}$ is an irreducible ideal of $\unlhd_{N \setminus \braces{I_1 \cup \dots \cup I_{i-1}} \setminus I_i}$.
        Thus, we can apply \Cref{lem:mergeconsecblocksfallsfalschrum} to the instance
        $(N \setminus \braces{I_1 \cup \dots \cup I_{i-1}}, \unlhd_{N \setminus \braces{I_1 \cup \dots \cup I_{i-1}}} ,c)$ with ideals $I_i$ and $I_{i+1}$.
        As $I_i \cupdot I_{i+1}$ is an ideal of $\unlhd_{N \setminus \braces{I_1 \cup \ldots \cup I_{i-1}}}$, it could have been a choice in iteration $i$ of the algorithm.
        Since $I_i$ was chosen to be inclusion-maximal among all minimal ideals w.r.t.~$\preceq$, we get from \Cref{lem:mergeconsecblocksfallsfalschrum}, that $I_i \preceq I_{i+1}$ has to hold.

    		By \cref{obs:algorithm:feasible}, we know that $S$ is feasible.
        Optimality of $S$ follows from \Cref{lem:iisoptimal}.
    %
    \end{proof}
\end{theorem}

If the set of jobs $N$ is not empty, also an inclusion-maximal minimal ideal in $\unlhd$ is non-empty.
Thus, \Cref{alg:minbudget:generic} terminates after at most $\card{N}$ many iterations.
Together with the previous theorem, the existence of schedules in increasing irreducible structure follows.

\begin{corollary}[Existence and Optimality of Schedules in Increasing Irreducible Structure]
\label{cor:incrirredexopt}
    Let $(N,\unlhd,c)$ be an instance of \minbudget.
    There exists a schedule in increasing irreducible structure and it is optimal for $(N,\unlhd,c)$.
\end{corollary}


Recall that for a schedule being in increasing irreducible structure with irreducible intervals $I_1,\dots,I_k$, we only require $I_1 \preceq I_2 \preceq \cdots \preceq I_k$.
Therefore $I_1$ is an ideal, but there might be another ideal $I$ with $I \prec I_1$ that is not scheduled contiguously in the schedule.
The following lemma shows that $I_1$ can be chosen to be minimal w.r.t.~$\preceq$.
It also shows that the first $l$ intervals of a schedule in increasing irreducible structure solve certain optimization problems.
This will be useful for our results in \Cref{sec:algorithms} to show that the schedules constructed there are again in increasing irreducible structure.

\begin{lemma}[Prefixes of Schedules in Increasing Irreducible Structure]
\label{lem:universal:prefix}
    Let $(N, \unlhd, c)$ be an instance of \minbudget.
    Let $S =  S_1 \oplus \ldots \oplus S_k$ be a schedule in increasing irreducible structure with corresponding irreducible intervals $I_1,\dots,I_k$.
    Let $B \in \Rbbp \cup \set{\infty}$.

    \begin{lemmaenum}
        \item \label{lem:universal:prefix:mincost}
        There is $l \in \set{0, \ldots, k}$ such that $I_1 \cup \ldots \cup I_l$ solves $\min \set{c(I): I \subseteq N \text{ ideal of } \unlhd \text{ with } b(I) \le B}$.
        \item \label{lem:universal:prefix:minreturn}
        There is $l \in \set{0, \ldots, k}$ such that $I_1 \cup \ldots \cup I_l$ solves $\min \set{r(I): I \subseteq N \text{ ideal of } \unlhd}$.
        \item \label{lem:irred:prefixminideal}
        There is $l \in \set{0, \ldots, k}$ such that $I_1 \cup \ldots \cup I_l$ is a minimal ideal of $\unlhd$ w.r.t.~$\preceq$.
    \end{lemmaenum}
    \begin{proof}
        \begin{lemmaenum}
        \item
        Choose $l \in \set{1, \ldots, k}$ such that $c(S_1 \oplus \ldots \oplus S_l)$ is minimal while $b(S_1 \oplus \ldots \oplus S_l) \le B$. If $b(S_1) > B$, set $l = 0$.
        We show that the corresponding set of jobs $L := I_1 \cup \ldots \cup I_l$ solves the minimization problem.
        First observe that $L$ is indeed an ideal of $\unlhd$ as $S$ is feasible and by definition of $l$ we have $b(L) \le B$.
        It remains to show that $L$ has minimal cost among all such ideals of $\unlhd$.

        Let $I \subseteq N$ be an ideal of $\unlhd$ with $b(I) \le B$ minimizing the cost $c(I)$ and let $S^*_I$ be an optimal schedule of $(I, \unlhd_I, c)$.
        Assume for a contradiction $c(I) < c(L)$.
        We construct a prefix $P$ of $S^*_I$ with high cost, yielding a contradiction to $b(I) \le B$.
%
				Let $P \subseteq S_I^*$ be an inclusion-minimal prefix of $S^*_I$ such that there is an $i > l$ with $c(I_i) < 0$ and $c(P \cap I_i) \ge b(I_i)$.

				We first discuss that such a prefix of $S^*_I$ exists.
        Since $S$ is in increasing irreducible structure and by the choice of $l$, it holds $c(S_j) = c(I_j) \le 0$ for all $j \in [l]$.
				Together with $I_j \cap I$ being an ideal of $\unlhd_{I_j}$ and \Cref{lem:irreducibleinterval:mincostideal}, this yields that $c(I_j \cap I) \geq c(I_j)$ for all $j \in [l]$.
        So $c(L \cap I) = \sum_{j \in [l]} c(I_j \cap I) \ge \sum_{j \in [l]} c(I_j) = c(L)$.
        The assumption $c(I) < c(L)$ implies $c(I \setminus L) < 0$.
        So there exists $i > l$ such that $c(I_i \cap I) < 0$.
        Since $I_i$ is irreducible this implies $c(I_i) < 0$, and by \Cref{lem:irreducibleinterval:ideal}, it holds $b(I_i \cap I) \ge b(I_i)$.
        As $S^*_I \cap I_i$ is a feasible schedule for $I_i \cap I$ it holds
        $ b(I \cap I_i) \le b \braces{S^*_I \cap I_i} = \max \set{c(P' \cap I_i) : P' \text{ prefix of } S^*_I}$.
        So there exist $i > l$ and a prefix $P$ of $S^*_I$ with $c(P \cap I_i) \geq b(I \cap I_i )\geq b(I_i)$ and $c(I_i) < 0$.

				Now let $P$ be inclusion-minimal such that there is $i > l$ with $c(I_i) < 0$ and $c(P \cap I_i) \ge b(I_i)$.
				As $P$ is a prefix of $S^*_I$, the jobs in $P \cap I_j$ are an ideal of $\unlhd_{I_j}$ for all $j \in [k]$.
				To lower bound the cost of $P$, we want to lower bound the cost of $P \cap I_j$ for all $j \in [k]$.
				For $j \in [l]$ we get from $c(I_j) < 0$ and \Cref{lem:irreducibleinterval:mincostideal} that $c(P \cap I_j) \ge \min\{0,c(I_j)\} = c(I_j)$.
				For $j > l$ and $c(I_j) \geq 0$, \cref{lem:irreducibleinterval:mincostideal} implies $c(P \cap I_j) \geq \min\{0,c(I_j)\} = 0$.
				Finally for $j > l$ with $j \not= i$ and $c(I_j) < 0$ the inclusion-minimality of $P$ implies $b(P \cap I_j) < b(I_j)$ (otherwise $P$ could have been chosen smaller).
				In particular the budget of the set of jobs in $P \cap I_j$ is less than $b(I_j)$ by \Cref{def:budgetreturninterval}.
				The irreducibility of $I_j$ and \cref{lem:irreducibleinterval:ideal} imply $c(P \cap I_j) \geq 0$.

			  For the cost of $P$, we obtain
        \begin{equation*}
            c(P) = \sum_{j \le l} c(P \cap I_j) + \sum_{j > l} c(P \cap I_j) \ge \sum_{j \le l} c(I_j) +c(P \cap I_i) \geq c(L) + b(I_i).
				\end{equation*}

				From the increasing irreducible structure of $S$, and $c(I_i) < 0$ (with $i > l$), we have $c(I_{l+1}) < 0$ and thus $b(I_i) \ge b(I_{l+1})$.
				From $c(I_{l+1}) = c(S_{l+1}) < 0$ we know that $S_1 \oplus \ldots \oplus S_{l+1}$ could have been a choice for $L$.
				But as we chose $l$ and not $l+1$ it holds $b(S_1 \oplus \ldots \oplus S_{l+1}) > B$. Together with $b(S_1 \oplus \ldots \oplus S_{l}) \le B$ we obtain from \Cref{obs:budgetreturncost:composition:budget} that $b(S_1 \oplus \ldots \oplus S_{l+1}) = c(S_1 \oplus \ldots \oplus S_l) + b(S_{l+1})$.

				Putting everything togehter we get:
				\[c(P) \ge c(L) + b(I_i) \geq c(L) + b(I_{l+1}) = c(S_1 \oplus \ldots \oplus S_l) + b(S_{l+1}) > B.\]
				Since $P$ is prefix of an optimal schedule $S^*_I$ for $I$ it holds $b(I) \ge c(P) > B$, which contradicts the choice of $I$.

        \item
        Let $l \in \set{1, \ldots, k}$ be maximal such that $c(S_l) < 0$. If there is no such $S_l$, set $l = 0$.
        Since $S$ is in increasing irreducible structure, $c(S_i) < 0$ for all $i \in [l]$ and $c(S_i) \ge 0$ for all $i > l$.
        We show that $I_1 \cup \ldots \cup I_l$ or $I_1 \cup \ldots \cup I_{l+1}$ has minimal return among all ideals of $\unlhd$.
        If $l = k$, we set $I_{l+1} = \emptyset$ in the following.

        First observe that both $I_1 \cup \ldots \cup I_l$ and $I_1 \cup \ldots \cup I_{l+1}$ are indeed ideals of $\unlhd$ as $S$ is feasible.
        For any ideal $I \subseteq N$ of $\unlhd$ we show that
        $r(I) \ge \min \set{r(S_1 \oplus \ldots \oplus S_l), r(S_1 \oplus \ldots \oplus S_{l+1}) } $, which yields the claim.

        Let $I \subseteq N$ be an ideal of $\unlhd$.
        Define the schedule $S^\prime := S^\prime_1 \oplus \ldots \oplus S^\prime_k$ for $(I,\unlhd_I,c)$ where, for every $i \in [k]$, $S^\prime_i$ is an optimal schedule of $I_i \cap I$.
        As $I$ is an ideal in $\unlhd$ and $I_i$ is an interval in $\unlhd$, we know that $I_i \cap I$ is an ideal of $\unlhd_{I_i}$ for every $i \in [k]$.
        As $S$ is feasible and the order of $I_i \cap I$ for all $i \in [k]$ is not changed in $S'$, also $S'$ is feasible for $(I,\unlhd_I,c)$.

        Applying \Cref{lem:irreducibleinterval:minreturnideal} to the irreducible interval $I_i$ and its ideal $I \cap I_i$, we get $r(I \cap I_i) \ge r(I_i)$.
        Since $S'_i$ is optimal for $I \cap I_i$ and $S_i$ is optimal for $I_i$, this implies $r(S^\prime_i) \ge r(S_i)$ for all $i \in [k]$.
        Further, \Cref{lem:irreducibleinterval:mincostideal} implies $c(I \cap I_i) \ge \min \set{0,c(I_i)}$ for all $i \in [k]$.
        So, by the choice of $l$, we get $ c(I \cap I_i) \ge c(I_i)$ for all $i \in [l]$.

        Since $S^\prime$ is feasible for $(I,\unlhd_I,c)$, we know $r(I) \geq r(S')$.
        By applying \Cref{obs:budgetreturncost:composition:return} multiple times, we obtain
        \begin{equation*}
        r(I) \geq r(S^\prime) = \min_{j \in [k]} \braces{r(S^\prime_j) + \sum_{i = j + 1}^k c(S^\prime_i) }.
        \end{equation*}

        Recall that $r(S'_i) \ge r(S_i)$ for all $i \in [k]$ and $c(S'_i) \ge c(S_i)$ for all $i \in [l]$. Further for all $i > l$, it holds $c(S_i) \ge 0$ and, thus again from \Cref{lem:irreducibleinterval:mincostideal}, $c(S'_i) \ge 0$. Therefore,

        \begin{equation*}
          r(I) \ge \min_{j \in [k]} \braces{r(S^\prime_j) + \sum_{i = j + 1}^k c(S^\prime_i) } \ge \min_{j \in [k]} \braces{ r(S_j) + \sum_{i = j + 1}^l c(S_i) }.
        \end{equation*}

        Applying \Cref{obs:budgetreturncost:composition:return} again multiple times yields
        \begin{equation*}
					 r(I) \geq  \min \set{
					 			r(S_1 \oplus \ldots \oplus S_l),
								\min_{j > l} r(S_j)
					 }.
        \end{equation*}

        Since $S$ is in increasing irreducible structure and by the choice of $l$, we get $r(S_{l+1}) = r(I_{l+1}) \le r(I_j) = r(S_j)$ for all $j > l$.
      	Note that $r(S_1 \oplus \ldots \oplus S_{l+1}) \leq r(S_{l+1})$ by \cref{obs:budgetreturncost:composition:return}. Thus,
      	\begin{equation*}
            r(I) \geq \min \set{r(S_1 \oplus \ldots \oplus S_l), r(S_{l+1}) } \geq \min \set{r(S_1 \oplus \ldots \oplus S_l), r(S_1 \oplus \ldots \oplus S_{l+1}) }.
       \end{equation*}

        \item
        Let $I \subseteq N$ be a minimal ideal of $\unlhd$ w.r.t.~$\preceq$.
        We distinguish between two cases.

        \emph{Case $c(I) < 0$:}
        By \cref{lem:universal:prefix:mincost}, there is $l \in \set{0, \ldots, k}$ such that the ideal $I_1 \cup \ldots \cup I_l$ solves
        $\min \set{c(J): J \subseteq N \text{ ideal of } \unlhd \text{ with } b(J) \le b(I)} \le c(I)$.
        As $I$ is minimal with negative cost, we have that $c(I_1 \cup \ldots \cup I_l) = c(I) < 0$ and
        $b(I_1 \cup \ldots \cup I_l) = b(I)$ and thus $I_1 \cup \ldots \cup I_l$ is a minimal ideal of $\unlhd$ w.r.t.~$\preceq$.

        \emph{Case $c(I) \ge 0$:}
        By \cref{lem:universal:prefix:minreturn}, there is $l \in \set{0, \ldots, k}$ such that the ideal $I_1 \cup \ldots \cup I_l$ solves $\min \set{r(J): J \subseteq N \text{ ideal of } \unlhd } \le r(I)$.
        As $I$ is a minimal ideal with non-negative cost, we obtain that $c(I_1 \cup \ldots \cup I_l) \ge 0$ and $r(I_1 \cup \ldots \cup I_l) = r(I)$ and thus $I_1 \cup \ldots \cup I_l$ is a minimal ideal of $\unlhd$ w.r.t.~$\preceq$. \qedhere
        \end{lemmaenum}
    \end{proof}
\end{lemma}

The third statement of the above lemma says that given a schedule in increasing irreducible structure, we can merge the first intervals $I_1,\dots,I_l$ to a minimal ideal w.r.t.~$\preceq$.
So from now on we may assume that the first interval of a schedule in increasing irreducible structure is in fact a minimal ideal w.r.t.~$\preceq$.
If not, we can merge the first $l$ intervals according to \cref{lem:irred:prefixminideal}.

\section{Algorithms for Special Instances}\label{sec:algorithms}

In the previous section, we presented \Cref{alg:minbudget:generic}, which computes an optimal feasible schedule for any instance $(N,\unlhd,c)$ of \minbudget.
However, the two main steps, choosing an ideal $I$ that is minimal w.r.t~$\preceq$, and computing an optimal schedule $S^*_I$ for $I$, may already be \NPhard.
In this section, we propose polynomial-time algorithms for special classes of partial orders.
The algorithms are based on the fact that schedules in increasing irreducible structure are optimal.

\subsection{Series-Parallel Partial Orders}\label{sec:seriesparallel}


In this subsection, we show that there is a polynomial-time algorithm for the \minbudget~problem for instances with precedence constraints represented by series-parallel partial orders.

We use the standard definition of series-parallel partial orders, see \cite{graphclasses}.
\begin{definition}[Series-Parallel Partial Orders]
  \emph{Series-parallel} partial orders are defined recursively as follows.
	\begin{definitionenum}
		\item A single element is a series-parallel partial order
		\item If $(N_1, \unlhd_1)$ and $(N_2, \unlhd_2)$ (with $N_1 \cap N_2 = \emptyset$) are two series-parallel partial orders, the \emph{parallel composition} $(N_1 \cupdot N_2, \unlhd_1 || \unlhd_2)$ is again a series-parallel partial order, where $\unlhd_1 || \unlhd_2 := \unlhd_1 \cupdot \unlhd_2$.
		\item If $(N_1, \unlhd_1)$ and $(N_2, \unlhd_2)$ (with $N_1 \cap N_2 = \emptyset$) are two series-parallel partial orders, the \emph{series composition} $(N_1 \cupdot N_2, \unlhd_1 * \unlhd_2)$ is again a series-parallel partial order, where $\unlhd_1 * \unlhd_2 := \unlhd_1 \cupdot \unlhd_2 \cupdot \set{(x,y) \colon x \in N_1, y \in N_2} $.
	\end{definitionenum}
\end{definition}

A series-parallel partial order can be represented by its binary decomposition tree, where the leaves are single elements and the internal nodes are the operators $||$ and $*$ (see \cite{Mohring1989}). Such a decomposition tree can be computed in polynomial time.

Our algorithm uses the decomposition tree of the partial order and computes recursively a schedule in increasing irreducible structure.
The recursion starts with single jobs and the schedule consisting of this job which is in increasing irreducible structure.
We show how to compute a schedule in increasing irreducible structure for parallel and series composition, given two schedules in increasing irreducible structure for the components.
By \Cref{lem:iisoptimal}, we know that the resulting schedule is optimal.


\paragraph{Parallel Composition}
For parallel compositions, we prove that a binary merge of the blocks of the schedules of the components w.r.t.~the cbr-preorder of the parallel composition yields the right schedule.

Observe that the cbr-preorder $\preceq$ of the parallel composition $(N_1 \cupdot N_2, \unlhd_1 || \unlhd_2, c_1 \cup c_2)$ restricted to subsets of $N_1$ or $N_2$ is the cbr-preorder of $(N_1, \unlhd_1, c_1)$ or $(N_2, \unlhd_2, c_2)$, respectively.
By $c_1 \cup c_2 = c$ we denote the costs on $N_1 \cup N_2$ where $c(j) := c_1(j)$ for jobs $j \in N_1$ and $c(j) := c_2(j)$ for jobs $j \in N_2$.

\begin{lemma}[Parallel Composition]
\label{lem:composition:parallel}
	Let $(N_1, \unlhd_1, c_1)$ and $(N_2, \unlhd_2,c_2)$ be two instances of \minbudget~with $N_1 \cap N_2 = \emptyset$.
    Let $S^1 = S^1_1 \oplus \ldots \oplus S^1_k$ and $S^2 = S^2_1 \oplus \ldots \oplus S^2_l$ be corresponding schedules in increasing irreducible structure with corresponding irreducible intervals $I^1_1,\dots,I^1_k$ and $I^2_1,\dots,I^2_l$, respectively.

    Sort $\set{I^1_1,\dots,I^1_k} \cup \set{I^2_1,\dots,I^2_l}$ in non-decreasing order with respect to the cbr-preorder of the parallel composition $(N_1 \cup N_2, \unlhd_1 || \unlhd_2, c_1 \cup c_2)$.
    Then scheduling the irreducible intervals with their optimal schedules in this order while keeping the relative ordering of $I^1_1,\dots,I^1_k$ and $I^2_1,\dots,I^2_l$ gives a feasible schedule in increasing irreducible structure for the parallel composition $(N_1 \cup N_2, \unlhd_1 || \unlhd_2, c_1 \cup c_2)$.
    \begin{proof}

    Note that from the comment following \Cref{def:irreducibility} of irreducible intervals, we have that the irreducible intervals $I^1_1, \ldots, I^1_k$ of $\unlhd_1$ and $I^2_1, \ldots, I^2_l$ of $\unlhd_2$ are also irreducible intervals of $\unlhd_1 || \unlhd_2$.

    When merging the two schedules $S^1$ and $S^2$ blockwise w.r.t.~$\preceq$, we maintain the (optimal) schedules $S^1_1, \ldots, S^1_k$ and $S^2_1, \ldots, S^2_l$ and their ordering and, thus, we obtain a feasible schedule of the parallel composition.
    By definition, this schedule is in line with $\preceq$. Hence the given schedule is in increasing irreducible structure.
    \end{proof}
\end{lemma}

\paragraph{Series Composition}
We show that the concatenation of the schedules of the two components yields a schedule of the series composition in increasing irreducible structure.
The intervals of the composition used in the schedule are exactly the irreducible intervals of the components except for possibly some of the intervals are merged.
The merging of intervals is necessary in order to establish that the resulting irreducible intervals are in increasing order w.r.t.\ $\preceq$.
The main property of irreducible interval that we use for this is stated in \cref{lem:mergeconsecblocksfallsfalschrum}.
Further, we obtain that the optimal schedule for the merged intervals are the corresponding parts of the concatenated schedule.

First, we observe that concatenating the two schedules in increasing irreducible structure is an optimal schedule for the series composition.

\begin{lemma}[Concatenation of Increasing Irreducible Schedules]
\label{lem:composition:series:concat}
Let $(N_1, \unlhd_1, c_1)$ and $(N_2, \unlhd_2,c_2)$ be two instances of \minbudget~with $N_1 \cap N_2 = \emptyset$.
Let $S^1$ and $S^2$ be corresponding schedules in increasing irreducible structure.

Then $S = S^1 \oplus S^2$ is an optimal schedule of $(N_1 \cup N_2, \unlhd_1 * \unlhd_2, c_1 \cup c_2)$.

\begin{proof}
  Let $S^1 = S^1_1 \oplus \ldots \oplus S^1_k$ and $S^2 = S^2_1 \oplus \ldots \oplus S^2_l$ with corresponding irreducible intervals $I^1_1,\dots,I^1_k$ of $\unlhd_1$ and $I^2_1,\dots,I^2_l$ of $\unlhd_2$, respectively.

Let $S^*$ be an optimal schedule of $(N_1 \cup N_2, \unlhd_1 * \unlhd_2, c_1 \cup c_2)$.
Recall from the comment to \Cref{def:irreducibility}, that the irreducible intervals $I^1_1, \ldots, I^1_k$ of $\unlhd_1$ and $I^2_1, \ldots, I^2_l$ of $\unlhd_2$ remain irreducible in $\unlhd_1 * \unlhd_2$.


As $S^*$ is feasible for $\unlhd_1 * \unlhd_2$, it is of the form $S_{N_1} \oplus S_{N_2}$ where $S_{N_1}$ and $S_{N_2}$ are feasible (but not necessarily optimal) schedules for $(N_1,\unlhd_1)$ and $(N_2,\unlhd_2)$.
Apply \Cref{lem:nonpreemption} to $S^*$ and the irreducible intervals $I^1_1, \ldots, I^1_k$ and $I^2_1, \ldots, I^2_l$.
We obtain a schedule $S'$ that schedules each interval $I^1_1,\dots,I^1_k$ and $I^2_1,\dots,I^2_l$ contiguously and in its respective optimal schedule $S^1_1,\dots,S^1_k$ and $S^2_1,\dots,S^2_l$.
The schedule satisfies $b(S') \leq b(S^*)$, i.e.~it is also optimal.
Note that by the construction of the prefix in \Cref{lem:nonpreemption} the intervals of the two components are not mixed.
We schedule the interval contiguously at a point of the schedule where some job of the interval was scheduled.
As we start with a feasible schedule, all jobs of $N_1$ remain before the jobs in $N_2$. So the resulting schedule is of the form $S' = S'_{N_1} \oplus S'_{N_2}$ where $S'_{N_1}$ and $S'_{N_2}$ schedule the intervals $I^1_1,\dots,I^1_k$ and $I^2_1,\dots,I^2_l$ contiguously and optimally, respectively.

Then applying \Cref{lem:consistency} independently to $S'_{N_1}$ and $S'_{N_2}$, we obtain the schedules $S^1$ and $S^2$, respectively.
Hence $b(S) \leq b(S') \leq b(S^*)$, which proves the claim.
\end{proof}

\end{lemma}

For recursive calls of the algorithm, however, optimality of the schedule does not suffice.
We need the schedules to be in increasing irreducible structure for merging in the parallel compositions.

The following lemma is key for the increasing irreducible structure for a series composition.
It states that there is a prefix of the two concatenated schedules which is the optimal schedule of a minimal ideal as well as the union of irreducible intervals from increasing irreducible structure.
In some sense, this extends \cref{lem:irred:prefixminideal}.

%
%
%

\begin{lemma}[Minimal Ideals in Series Compositions]\label{lem:minidealofseriesconcat}
    Let $(N_1, \unlhd_1, c_1)$ and $(N_2, \unlhd_2,c_2)$ be two instances of \minbudget~with $N_1 \cap N_2 = \emptyset$.
    Let $S^1 = S^1_1 \oplus \ldots \oplus S^1_k$ and $S^2 = S^2_1 \oplus \ldots \oplus S^2_l$ be corresponding schedules in increasing irreducible structure with corresponding irreducible intervals $I^1_1,\dots,I^1_k$ and $I^2_1,\dots,I^2_l$, respectively.

    There are $p \in \set{0, \ldots, k}$ and $q \in \set{0, \ldots, l}$ such that $I^1_1 \cup \dots \cup I^1_p$ or $N_1 \cup I^2_1 \cup \dots \cup I^2_q$ is a minimal ideal of $\unlhd_1 * \unlhd_2$ w.r.t.\ $\preceq$.

    \begin{proof}
      Let $I \subseteq N_1 \cup N_2$ be any minimal ideal of $\unlhd_1 * \unlhd_2$ w.r.t.\ $\preceq$.
      If $I \subseteq N_1$, we know from \cref{lem:irred:prefixminideal} that there is $p \in \set{0, \ldots, k}$ such that $I^1_1 \cup \dots \cup I^1_p$ is a minimal ideal in $N_1$.
      \\
      Otherwise, write $I = N_1 \cupdot J_2$.
      We show that the equations from \Cref{obs:budgetreturnconcat} hold for the job sets (and not only for particular schedules).
      Using the fact that any feasible schedule of $N_1 \cupdot J_2$ has to schedule $N_1$ completely before $J_2$, we obtain
      $b(N_1 \cupdot J_2) = \max \set{b(N_1), c(N_1) + b (J_2)}$ and $r(N_1 \cupdot J_2) = \min \set{r(N_1) + c(J_2), r(J_2)}$.
      Thus, for $I = N_1 \cupdot J_2$, we get $b(I) = \max \set{b(N_1),c(N_1) + b(J_2)} \ge b(N_1)$.
      Depending on the cost of $I$, we distinguish two cases.
      \begin{itemize}[align=left,leftmargin=*,font=\itshape]
      \item[Case $c(I) < 0$:]
          Since $I$ is minimal w.r.t.~$\preceq$ and has negative cost, we get from \Cref{def:cbrpreorder} that $J_2$ is minimizer of the following expression
          \begin{align*}
          &\min \set{ r(N_1 \cupdot J): J \subseteq N_2 \text{ ideal of } \unlhd_2 \text{ with } c(N_1 \cupdot J) < 0 \text{ and } b(N_1 \cupdot J) \le b(I) } \\
          = & \min \set{ c(N_1) + c(J) - b(N_1 \cupdot J):
                J \subseteq N_2 \text{ ideal of } \unlhd_2 \text{ with } c(N_1 \cupdot J) < 0 \text{ and }
                 b(N_1 \cupdot J) \le b(I) }.
          \end{align*}
          From the restriction $ b(N_1 \cupdot J) \le b(I) $ and as $I$ is minimal w.r.t.~$\preceq$, we get $b(N_1 \cupdot J) = b(I) $.
          Since additionally $b(N_1) \le b(I)$ and $b(N_1 \cupdot J) = \max \set{ b(N_1), c(N_1) + b(J) } $, we obtain that $J_2$ is minimizer of
          \[  \min \set{ c(J): J \subseteq N_2 \text{ ideal of } \unlhd_2 \text{ with } c(N_1)+ b(J)  \le b(I)} + c(N_1) - b(I) . \]
          Now from \Cref{lem:universal:prefix:mincost}, we know that there is $q \in \set{0, \ldots, l}$ such that $I^2_1 \cup \ldots \cup I^2_q$ solves this optimization problem.
          Hence, $N_1 \cup I^2_1 \cup \ldots \cup I^2_q$ is a minimal ideal of $\unlhd_1 * \unlhd_2$.
      \item[Case $c(I) \ge 0$:]
          Since $I$ is minimal w.r.t.~$\preceq$ and has non-negative cost we get from \Cref{def:cbrpreorder} that $J_2$ is minimizer of the following expression
          \begin{align*}
          &\min \set{ r(N_1 \cupdot J): J \subseteq N_2 \text{ ideal of } \unlhd_2 } \\
          = & \min \set{ \min \set{r(N_1) + c(J), r(J)}: J \subseteq N_2 \text{ ideal of } \unlhd_2 } \\
          = &\min
          \left\{\!\begin{aligned}
          r(N_1) + \min \set{ c(J): J \subseteq N_2 \text{ ideal of } \unlhd_2 \text{ with } b(J) \le -r(N_1) },\\
            \min \set{ r(J): J \subseteq N_2 \text{ ideal of } \unlhd_2 } \end{aligned}\right\}.
          \end{align*}
          From \Cref{lem:universal:prefix:mincost,lem:universal:prefix:minreturn}, we know that there are $q_1,q_2 \in \set{0, \ldots, l}$ such that $I^2_1 \cup \ldots \cup I^2_{q_1}$ and $I^2_1 \cup \ldots \cup I^2_{q_2}$
          solve the two inner minimization problems.
          Hence, $N_1 \cup I^2_1 \cup \ldots \cup I^2_{q_1}$ or $N_1 \cup I^2_1 \cup \ldots \cup I^2_{q_2}$ is a minimal ideal of $\unlhd_1 * \unlhd_2$.
      \end{itemize}
      In both cases, there is $q \in \set{0, \ldots, l}$ such that $N_1 \cup I^2_1 \cup \dots \cup l^2_q$ is a minimal ideal of $\unlhd_1 * \unlhd_2$.
    \end{proof}
\end{lemma}

We now can combine the above lemmas in order to prove that the concatenation of the schedules of two components is indeed a schedule in increasing irreducible structure for their series composition.
The proof of the following theorem is constructive.
It shows that, similar to \cref{alg:minbudget:generic}, the concatenated schedule can be divided into irreducible intervals by iteratively finding prefixes that are minimal w.r.t.\ $\preceq$.

\begin{theorem}[Series Composition]
\label{thm:composition:series}
    Let $(N_1, \unlhd_1, c_1)$ and $(N_2, \unlhd_2,c_2)$ be two instances of \minbudget~with $N_1 \cap N_2 = \emptyset$.
    Let $S^1$ and $S^2$ be schedules in increasing irreducible structure for $(N_1, \unlhd_1, c_1)$ and $(N_2, \unlhd_2,c_2)$, respectively.

    Then $S^1 \oplus S^2$ is a feasible schedule in increasing irreducible structure of the series composition $(N_1 \cup N_2, \unlhd_1 * \unlhd_2, c_1 \cup c_2)$.

   \begin{proof}
        Let $S^1 = S^1_1 \oplus \ldots \oplus S^1_k$ and $S^2 = S^2_1 \oplus \ldots \oplus S^2_l$ with corresponding irreducible intervals $I^1_1,\dots,I^1_k$ and $I^2_1,\dots,I^2_l$.
        We show that $S^1 \oplus S^2$ is in increasing irreducible structure by exhibiting the corresponding irreducible intervals.

        We use an induction on $\card{N_1} + \card{N_2}$.
        If $N_1 \cup N_2 = \emptyset$, the statement is trivial.
        Assume the statement is true for $m \in \Nbb$ and $\card{N_1} + \card{N_2} < m$.
        We show that $S_1 \oplus S_2$ is in increasing irreducible structure if $\card{N_1} + \card{N_2} = m$.

        We define a minimal ideal $J$ of $\unlhd_1 * \unlhd_2$ that is a prefix of $S_1 \oplus S_2$ and contains as many of the intervals $I^1_1, \ldots, I^1_k, I^2_1, \ldots, I^2_l$ as possible.
        Applying \Cref{lem:minidealofseriesconcat} to $S^1$ and $S^2$, we have that there is
        $p \in \set{0, \ldots, k}$ such that $I^1_1 \cup \dots \cup I^1_p$ is a minimal ideal of $\unlhd_1 * \unlhd_2$ w.r.t.\ $\preceq$,
        or there is $q \in \set{0, \ldots, l}$ such that $N_1 \cup I^2_1 \cup \dots \cup I^2_q$ is a minimal ideal of $\unlhd_1 * \unlhd_2$ w.r.t.\ $\preceq$.
        If there is $q \in \set{0, \ldots, l}$ such that $N_1 \cup I^2_1 \cup \dots \cup I^2_q$ is a minimal ideal of $\unlhd_1 * \unlhd_2$, pick $q$ maximal and set $J := N_1 \cup I^2_1 \cup \dots \cup I^2_q$ to be a minimal ideal of $\unlhd_1 * \unlhd_2$.
        Otherwise, we know there has to be $p \in \set{0, \ldots, k}$ such that $I^1_1 \cup \dots \cup I^1_p$ is a minimal ideal of $\unlhd_1 * \unlhd_2$.
        Pick $p$ maximal and set $J := I^1_1 \cup \dots \cup I^1_p$ to be a minimal ideal of $\unlhd_1 * \unlhd_2$.
        If $\emptyset$ is a minimal ideal w.r.t.~$\preceq$, then the instance has only jobs with non-negative cost.
        The maximal choice of $J$ yields $J \ne \emptyset$, since $N_1 \cup N_2 \ne \emptyset$.
        \\
        $S_1 \cap J$ and $S_2 \cap J$ are in increasing irreducible structure with corresponding irreducible intervals $I^1_1, \ldots, I^1_k$ and $I^2_1, \ldots, I^2_l$ that are contained in $J$.
        By \cref{lem:composition:series:concat}, we know that the prefix $(S_1 \oplus S_2) \cap J = (S_1 \cap J) \oplus (S_2 \cap J)$ is scheduled optimally in $S_1 \oplus S_2$.
        Also, $S_1 \setminus J$ and $S_2 \setminus J$ are in increasing irreducible structure with corresponding irreducible intervals $I^1_1, \ldots, I^1_k$ and $I^2_1, \ldots, I^2_l$ which are not contained in $J$.
        Therefore, the induction hypothesis implies that the suffix $(S_1 \oplus S_2) \setminus J = (S_1 \setminus J) \oplus (S_2 \setminus J)$ is in increasing irreducible structure again.
        By applying \Cref{lem:minidealofseriesconcat} to $(N_1 \setminus J, \unlhd_{N_1 \setminus J}, c )$, $(N_2 \setminus J, \unlhd_{N_2 \setminus J}, c )$ with $S_1 \setminus J$ and $S_2 \setminus J$, we know that $\unlhd_{(N_1 \cup N_2) \setminus J}$
        has a minimal ideal $L$ given by the first intervals of $I^1_1, \ldots, I^1_k, I^2_1, \ldots, I^2_l$ which are not contained in $J$.
        As $J$ and $L$ are minimal w.r.t.~$\preceq$, they are irreducible.
        Since $J \cup L$ was a feasible choice in the definition of $J$ and $J$ was chosen to be inclusion-maximal, \cref{lem:mergeconsecblocksfallsfalschrum} yields that $J \preceq L$.
        In total, $S_1 \oplus S_2 = \braces{(S_1 \oplus S_2) \cap J} \oplus \braces{(S_1 \oplus S_2) \setminus J}$ is in increasing irreducible structure again.
   \end{proof}

\end{theorem}

\begin{corollary}[Series-Parallel Partial Orders]
\label{cor:seriesparallel:computation}
    For an instance $(N, \unlhd, c)$ of \minbudget, where $\unlhd$ is a series-parallel partial order,
    a schedule in increasing irreducible structure can be computed in polynomial time.
    \begin{proof}
      Compute the decomposition tree of $\unlhd$ and compute a schedule in increasing irreducible structure recursively beginning at the leaves of the tree.
      Note that the cost, budget, and return of a subschedule can be computed efficiently based on their defintions.
      The cost, budget, and return of the corresponding irreducible intervals of an increasing irreducible structure can also be determined, as the corresponding parts of the schedule are optimal.
      Therefore, the comparisons needed for the sorting w.r.t.~$\preceq$ in \Cref{lem:composition:parallel} can be done in polynomial time.
      Similarly, the prefixes of \Cref{lem:minidealofseriesconcat} that are needed in \Cref{thm:composition:series} can be found in polynomial time.
    \end{proof}
\end{corollary}

\subsection{Convex Bipartite Partial Orders}\label{sec:convex}


In this section, we propose an algorithm to solve \minbudget~ for instances with convex bipartite partial orders.
To simplify notation, let $N = N^+ \cupdot N^-$ be the job set with $N^+ = \set{ j \in N : c_j \geq 0}$ and $N^- = \set{j \in N : c_j < 0}$ and $\unlhd \subseteq N^+ \times N^-$ a bipartite partial order.
Further let $\mathcal{P}(j) \subseteq N^+$ be the set of predecessors of $j \in N^-$, and $\mathcal{S}(i) \subseteq N^-$ the set of successors of $i \in N^+$.

\begin{definition}[Convex Bipartite Partial Order]
    \label{def:convex}
    Let $N^+$ and $N^-$ be two job sets
    and let $\unlhd \subseteq N^+ \times N^- $ be a bipartite partial order on $N^+ \cupdot N^-$.

    We call $\unlhd$ \emph{convex in $N^+$} if there exists a linear order $<$ on $N^+$ such that for every $j \in N^-$ the set of predecessors $\mathcal{P}(j)$ is an interval w.r.t.~$<$.

    We call $\unlhd$ \emph{convex in $N^-$} if there exists a linear order $<$ on $N^-$ such that, for every $i \in N^+$, the set of successors $\mathcal{S}(i)$ is an interval w.r.t.~$<$.

    We call $\unlhd$ \emph{convex} if it is convex in $N^+$ or convex in $N^-$.
\end{definition}

We present an algorithm that solves instances where the precedence constraints are bipartite and convex in $N^-$, and use a simple observation to solve instances that are convex in $N^+$.
For a partial order $\unlhd$ we use the notation $\unlhd^{-1} = \set{(i,j) : j \unlhd i}$ for the dual partial order of $\unlhd$ (see \cite{Mohring1989}).

\begin{lemma}[Reverse Instance]
\label{lem:reverseinstance}

Let $(N, \unlhd, c)$ be an instance of \minbudget.
Define the \emph{reverse instance} as $\braces{N,\unlhd^{-1},-c}$.

Then $S$ is a feasible schedule for $(N, \unlhd, c)$ with budget $b$ and return $r$ if and only if $S^{-1}$ is a feasible schedule for $\braces{N,\unlhd^{-1},-c}$ with budget $\overline{b} = - r$ and return $\overline{r} = - b$.

\begin{proof}
Note that $S$ is feasible for $(N,\unlhd,c)$ if and only if $S^{-1}$ is feasible for $ \braces{N,\unlhd^{-1},-c}$.
Let $\overline{c} = -c$.
The relation between budget and return of the respective schedules follows from the observation that $c(S) = c(N) = - \overline{c}(N) = -\overline{c} \braces{S^{-1} }$ is constant.
\end{proof}

\end{lemma}

Note that $(N,\unlhd,c)$ is convex in $N^+$, if and only if the reverse instance $ \braces{N, \unlhd^{-1},-c}$ is convex in $N^-$.
Hence, in order to solve an instance that is convex in $N^+$, we construct its reverse instance, use our algorithm for instances that are convex in $N^-$, and then transform the schedule accordingly (\cref{lem:reverseinstance}).
So let $\braces{N^+ \cup N^-,\unlhd,c}$ be an instance such that $\unlhd$ is bipartite and convex in $N^-$.
Key ingredient to the algorithm are two observations.

First, there is an optimal schedule that starts with a job $j \in N^-$ together with its predecessors $\mathcal{P}(j)$, and then schedules the jobs in $N_{-j} := N \setminus (\mathcal{P}(j) \cup \{j\})$ afterwards.
Second, if we delete $\mathcal{P}(j) \cup \{j\}$ from the instance, then $\braces{N_{-j}, \unlhd_{N_{-j}},c}$ decomposes into smaller instances $\braces{L, \unlhd_L, c}$ and $\braces{R, \unlhd_R,c}$ with $L \cupdot R = N_{-j}$.
For these instances we have $L = L^+ \cupdot L^-$ and $\unlhd_L$ is convex in $L^-$ and $R = R^+ \cupdot R^-$ and $\unlhd_R$ is convex in $R^-$.
We exploit this decomposition into smaller instances by using a dynamic program.
Informally, we iterate over all $j \in N^-$, fix $\mathcal{P}(j) \cup \{j\}$ to be the first jobs of the schedule, and schedule the instances $(L, \unlhd_L,c)$ and $(R, \unlhd_R,c)$ in parallel after $j$ using \cref{lem:composition:parallel}.
The respective schedules in increasing irreducible structure for $(L, \unlhd_L,c)$ and $(R, \unlhd_R,c)$ have been computed in earlier iterations of the dynamic program.

In order to pick the optimal schedule in the end, we make use of the first observation in \cref{lem:guessing}.
The idea can be summarized as follows.
If we correctly guess the first job $j \in N^-$ and its predecessors $\mathcal{P}(j)$ of an optimal schedule, then any optimal solution of $(N, \unlhd_j,c)$, where $\unlhd_j$ is defined as the series composition
$\braces{\mathcal{P}(j) \cup \{j\},\unlhd_{\mathcal{P}(j) \cup \{j\}}} * \braces{N_{-j}, \unlhd_{N_{-j}}}$, is an optimal solution of $(N, \unlhd_j,c)$.
In particular, any schedule in increasing irreducible structure for $(N,\unlhd_j,c)$ is also in increasing irreducible structure for $(N,\unlhd,c)$ and therefore optimal.
\Cref{lem:guessing} characterizes this first job.

\begin{lemma}[Guessing the first negative job]
\label{lem:guessing}
    Let $(N, \unlhd, c)$ be an instance of \minbudget~and $\preceq$ the corresponding cbr-preorder.
    Set $N^- := \set{j \in N: c_j < 0}$ and assume $N^- \ne \emptyset$.

    For $j \in N^-$ let $J_{j}$ be a minimial ideal of $\braces{N, \unlhd_j , c}$ w.r.t.~$\preceq$.

    Choose $j^* \in N^-$ such that the corresponding ideal $J_{j^*}$ is minimal w.r.t.~$\preceq$ among all ideals $J_{j}$, i.e.,~choose $j^*$ so that $J_{j^*}$ is a minimizer of
    \[  {\textstyle\min_\preceq} \set{J_{j} \colon j \in N^-}.
    \]

    Then any schedule in increasing irreducible structure for $\braces{N, \unlhd_{j^*} , c}$
    is a schedule in increasing irreducible structure for $(N, \unlhd, c)$.
    \begin{proof}
    Let $S = S_1 \oplus \ldots \oplus S_k$ be a schedule of $(N, \unlhd_{j^*}, c)$ in increasing irreducible structure and $I_1, \ldots, I_k$ be the corresponding irreducible intervals of $\unlhd_{j^*}$ such that $I_1$ is a minimal ideal of $\unlhd_{j^*}$ w.r.t.~$\preceq$ (by \Cref{lem:irred:prefixminideal}).
    $S$ is a feasible schedule of $(N, \unlhd, c)$ as $\unlhd_{j^*}$ is an extension of $\unlhd$.
    Also from $\unlhd_{j^*}$ being an extension of $\unlhd$ we have that ideals in $\unlhd_{j^*}$ are ideals in $\unlhd$ and the same holds for intervals.

    Further, we know that $\braces{\unlhd_{j^*}}_{N_{-j^*}} = \unlhd_{N_{-j^*}}$.
    Let $I$ be an interval of $\unlhd_{j^*}$ that does not intersect $\mathcal{P}(j^*) \cup \set{j^*}$.
    Note that ideals of $I$ in $\unlhd_{j^*}$ are exactly the same ideals in $\unlhd$, and thus $I$ is irreducible in $\unlhd_{j^*}$ if and only if it is irreducible in $\unlhd$.
    The first interval, i.e.,~$I_1$, contains $\mathcal{P}(j^*) \cup \set{j^*}$ as it is an ideal of the series composition $\unlhd_{j^*}$ and minimal w.r.t.~$\preceq$.

    Thus all intervals $I_2, \ldots, I_k$ are also irreducible in $\unlhd$.

    It remains to show that $I_1$ is an irreducible ideal in $\unlhd$.
    Recall that a minimal ideal is per definition irreducible. 
    As every ideal of $\unlhd_{j^*}$ is also an ideal of $\unlhd$, it remains to show that $I_1$ is minimal w.r.t.~$\preceq$ among all ideals of $\unlhd$.
    Let $I$ be a minimal ideal of $\unlhd$ w.r.t.~$\preceq$. 
    From $N^- \neq \emptyset$ and $I$ being minimal, we know that it contains a job $j \in N^-$ together with all its predecessors $\mathcal{P}(j)$.
    Thus, $I$ is an ideal in $\unlhd_{j}$ and so $ J_{j} \preceq I$ by the choice of $J_{j}$.
    As $I_1$ is minimal w.r.t.~$\preceq$ in $\unlhd_{j^*}$, we obtain $I_1 \preceq J_{j^*} \preceq J_{j} \preceq I$ by the choice of $J_{j^*}$.

    So, we showed that the irreducible intervals $I_1, \ldots, I_k$ of $\unlhd_{j^*}$ are also irreducible intervals in $\unlhd$, and, thus, the schedule $S$ is in increasing irreducible structure for $(N, \unlhd, c)$.
    \end{proof}

\end{lemma}

We now propose a dynamic programm (\Cref{alg:convex}) for instances $\braces{N^- \cupdot N^+, \unlhd, c}$ where $\unlhd$ is bipartite and convex in $N^-$ with corresponding linear order $<$.
Recall that there are schedules in increasing irreducible structure that start with a negative job $j \in N^-$ and its predecessors $\mathcal{P}(j)$.
Due to \Cref{lem:guessing}, it suffices to compute increasing irreducible schedules for the instances $(N, \unlhd_{j},c)$ for all $j \in N^-$, and pick the one with minimal first interval w.r.t.~the cbr-preorder.

The algorithm iteratively computes a schedule in increasing irreducible structure for the subinstance $\braces{I^- \cup I^+,\unlhd_{I^- \cup I^+},c}$ for all intervals $I^- \subseteq N^-$ w.r.t.~$<$, and a subset of their predecessors $I^+ = \set{i \in N^+ : \mathcal{S}(i) \subseteq I^-}$.
The idea is that all jobs in $N \setminus (I^- \cup I^+)$ are already fixed to be scheduled at the beginning of the schedule, and we want to compute an optimal schedule for the other jobs.
For every job $j \in I^-$, we consider the schedule where $j$ is scheduled after its predecessors in $I^+$ at the beginning of the schedule.
Note that the schedule $S_j$ of $j$ and its predecessors in $I^+$ (\Cref{alg:convex:jwithprecs}) is trivially in increasing irreducible structure.
If we remove $\set{j} \cup \mathcal{P}(j)$ from consideration, the instance decomposes into strictly smaller instances that are also convex.
The corresponding optimal schedules for these were computed in earlier iterations of the dynamic program, and so we can merge their schedules similar to \Cref{lem:composition:parallel} to an optimal schedule in \Cref{alg:convex:leftrightschedule}.
Finally we concatenate both schedules in \Cref{alg:convex:concatenation}, and derive a schedule in increasing irreducible structure for $(I^- \cup I^+, (\unlhd_{j})_{I^- \cup I^+},c)$  (\Cref{thm:composition:series}).
After we computed the schedules for all $j \in I^-$, we choose the best according to \Cref{lem:guessing} in \Cref{alg:convex:choosebest} to be the schedule for this subinstance.

\begin{algorithm}[h]
	\setstretch{1.25}
	\medskip
	\KwInput{$\braces{N^- \cupdot N^+, \unlhd, c}$ instance of \minbudget~where $\unlhd$ is convex in $N^-$ and respective order $<$ on $N^-$}
	\KwOutput{schedule of $\braces{N^- \cupdot N^+, \unlhd, c}$ in increasing irreducible structure}
	\medskip
    $S_{\emptyset} \leftarrow$ empty schedule\;
    \For{$k = 1, \ldots, \card{N^-}$}{
        \For{interval $I^- \subseteq N^-$ w.r.t.\ $<$ with $\card{I^-} = k$}{
            $I^+ \leftarrow \set{i \in N^+: \mathcal{S}(i) \subseteq I^-}$ \;
            \For{$j \in I^-$}{
                $L^-_j \leftarrow \set{i \in I^-: i < j}$, $R^-_j \leftarrow \set{i \in I^-: j < i}$, $\mathcal{P}^+_j \leftarrow \mathcal{P}(j) \cap I^+ $ \;

                $ S_j \leftarrow $ arbitrary feasible schedule of $\mathcal{P}^+_j \cup \set{j}$ \label{alg:convex:jwithprecs}\;
                $ S_{-j} \leftarrow \text{parallel composition of } S_{L^-_j} \text{ and } S_{R^-_j}$\label{alg:convex:leftrightschedule}
                \tcp*{\Cref{lem:composition:parallel}}

                $S'_j \leftarrow S_j \oplus S_{-j}$ \label{alg:convex:concatenation}\tcp*{in increasing irreducible structure}
                Compute the irreducible intervals $I^1_j, \ldots, I^k_j$ of $S'_j$ such that $I^1_j$ is a minimal ideal of $\unlhd_{j}$ w.r.t.~$\preceq$\label{alg:convex:computeintervals} \tcp*{\Cref{lem:minidealofseriesconcat,thm:composition:series}}
            }
            Let $j^* \in I^-$ be such that $I^1_{j^*}$ is minimizer of ${\textstyle\min_\preceq} \set{I^1_{j} \colon j \in I^-}$ \label{alg:convex:choosebest} \;
            $S_{I^-} \leftarrow S'_{j^*}$\label{alg:convex:intervalschedule} \;
        }
    }
    \Return $S_{N^-}$ \;
	\medskip
	\caption{Dynamic Program for Convex Bipartite Partial Orders}
	\label{alg:convex}
\end{algorithm}

\begin{theorem}[Correctness of \Cref{alg:convex}]
    \label{thm:convex}
    Let $(N, \unlhd, c)$ be an instance of \minbudget, where $N = N^- \cupdot N^+$ and $\unlhd$ is bipartite and convex in $N^-$. Let $<$ be a corresponding order on $N^-$.

    Then \Cref{alg:convex} computes a schedule for $(N, \unlhd, c)$ in increasing irreducible structure in polynomial time.
    \begin{proof}
        We prove that, for every interval $I^-$ of $N^-$ w.r.t.~$<$, the computed schedule $S_{I^-}$ is a schedule for $\braces{ I^+ \cup I^-, \unlhd_{I^+ \cup I^-} , c }$ in increasing irreducible structure.
        The correctness of the algorithm follows, as $I^+ = N^+$ in the last iteration (where $I^- = N^-$) and, hence, $S_{N^-}$ is a schedule for $\braces{N^+ \cupdot N^-, \unlhd, c}$ in increasing irreducible structure.

        We use induction on $\card{I^-}$.
        The statement is true for $\card{I^-} = 0$ and thus $I^- = \emptyset$ due to the initialization step of the algorithm.

        Fix $k \in \set{1, \ldots, \card{N^-}}$ and assume that the claim holds for every $I^- \subseteq N^-$ with $\card{I^-} < k$.
        Now, let $I^- \subseteq N^-$ with $\card{I^-} = k$ and consider the iteration in which $S'_j$ for a $j \in I^-$ was determined.
        Let $L^-_j, R^-_j$, and $\mathcal{P}^+_j$ be the sets as evaluated in that iteration.
        As $L^-_j,R^-_j \subseteq I^-$ we have $\card{L^-_j}, \card{R^-_j} < k$.
        We set $L^+_j := \set{i \in N^+: \mathcal{S}(i) \subseteq L^-_j} $ and $R^+_j := \set{i \in N^+: \mathcal{S}(i) \subseteq R^-_j}$ to be the corresponding $I^+$, respectively. $L^-_j,R^-_j \subseteq I^-$ directly implies $L^+_j, R^+_j \subseteq I^+$.

        Since $\unlhd$ is convex in $N^-$ the following holds: if $\mathcal{S}(i) \cap L^-_j \ne \emptyset$ and $\mathcal{S}(i) \cap R^-_j \ne \emptyset$ then $j \in \mathcal{S}(i)$ for any job $i \in N^+$.
        Therefore, the set $I^- \cup I^+$ can be partitioned into $\braces{ L^+_j \cup L^-_j } \cupdot \braces{ \mathcal{P}^+_j \cup \set{j} } \cupdot \braces{ R^+_j \cup R^-_j }$.
        Note $\braces{L^+_j \cup L^-_j, \unlhd_{L^+_j \cup L^-_j}} $ and $\braces{R^+_j \cup R^-_j, \unlhd_{R^+_j \cup R^-_j}} $ are convex in $L^-_j$ and $R^-_j$, respectively.
        Thus the schedules $S_{L^-_j}$ and $S_{R^-_j}$ were computed in previous iterations and are in increasing irreducible structure (induction hypothesis).

        Note that $S_j$ in \Cref{alg:convex:jwithprecs} is in increasing irreducible structure.
        By induction hypothesis and \Cref{lem:composition:parallel}, $S_{-j}$ is in increasing irreducible structure.
        Using \Cref{thm:composition:series}, it follows that $S'_j$ is in increasing irreducible structure.
        Thus $S'_j$ indeed is a feasible schedule for the series and parallel composition
        $\braces{\mathcal{P}^+_j \cup \set{j}, \unlhd_{\mathcal{P}^+_j \cup \set{j}}}
        * \braces{
            \braces{L^+_j \cup L^-_j, \unlhd_{L^+_j \cup L^-_j}}
            \ \| \
            \braces{R^+_j \cup R^-_j, \unlhd_{R^+_j \cup R^-_j}}
               }$.
        By \Cref{thm:composition:series,cor:seriesparallel:computation}, we can compute the irreducible intervals in \Cref{alg:convex:computeintervals} in polynomial time.
        $S_{I^-}$ can be computed efficiently from $S'_j$ by comparing the $k$ prefixes $I^1_j$ for $j \in I^-$.

        For every $j \in I^-$, we computed the feasible schedule $S'_j$ in increasing irreducible structure for the convex instance
        $\braces{I^+ \cup I^-, \unlhd_{j}}$.
        By the choice of $j^*$ in \Cref{alg:convex:choosebest} and \Cref{lem:guessing}, $S_{I^-}$ in \Cref{alg:convex:intervalschedule} indeed is a schedule for $\braces{ I^+ \cup I^-, \unlhd_{I^+ \cup I^-} }$ in increasing irreducible structure.
        Since the number of intervals in $N^-$ is quadratic in $\card{N^-}$, the number of iterations of the algorithm is polynomial.
    \end{proof}
\end{theorem}

\begin{corollary}[Convex Bipartite Partial Orders]
    For an instance $(N, \unlhd, c)$ of \minbudget, where $\unlhd$ is a convex bipartite partial order,
    a schedule in increasing irreducible structure can be computed in polynomial time.
    \begin{proof}
        The statement is implied by \Cref{thm:convex,lem:reverseinstance}.
    \end{proof}
\end{corollary}


\section{Conclusion and Future Work}\label{sec:conclusion}

We show that there exists an optimal solution of \minbudget~that is in increasing irreducible structure.
Further, we give polynomial time algorithms for series-parallel as well as for convex bipartite partial orders for which a joint generalization exists---two-dimensional partial orders.
Thus, this class is a natural candidate for further studies.
Another interesting direction, is to identify more classes of partial orders where the increasing irreducible structure can be exploited to design polynomial-time algorithms.
Due to the inapproximability in the general case, one might also consider different cost functions, e.g.,~such that $c(J) > 0$ for all non-empty ideals $J \subseteq N$, for which (constant) approximation algorithms might exist.

    \bibliography{minbudget}{}
    \bibliographystyle{amsplain}

\end{document}